\documentclass[
    ,final            ]
  {aipproc}

\pdfoutput=1
\usepackage{deluxetable}
\def\lesssim{\mathrel{\hbox{\rlap{\hbox{\lower4pt\hbox{$\sim$}}}\hbox{$<$}}}}
\def\gtrsim{\mathrel{\hbox{\rlap{\hbox{\lower4pt\hbox{$\sim$}}}\hbox{$>$}}}}
\def\negsp{\!\!\!\!\!\!\!\!\!}

\layoutstyle{8x11double}

\begin{document}

\newcommand{\edot}{\dot{E}}
\newcommand{\pdot}{\dot{P}}
\newcommand{\lpwn}{L_{\rm pwn}}
\newcommand{\lpsr}{L_{\rm psr}}
\newcommand{\etapwn}{\eta_{\rm pwn}}
\newcommand{\etapsr}{\eta_{\rm psr}}
\newcommand{\chan}{{\sl Chandra}\/}
\newcommand{\be}{\begin{equation}}
\newcommand{\ee}{\end{equation}}
\newcommand{\xmm}{{\sl XMM-Newton\/}}

\title{Pulsar Wind Nebulae in the Chandra Era}

\classification{97.60.Gb; 98.38.Mz; 98.38.-j; 98.70.Qy; 97.60.Jd}
\keywords      {Pulsar Wind Nebulae; Pulsars; Supernova Remnants; Neutron Stars}

\author{O. Kargaltsev}{
  address={Pennsylvania State University, 525 Davey Lab., University Park,
PA 16802, USA}
}

\author{G. G. Pavlov}{
  address={Pennsylvania State University, 525 Davey Lab., University Park,
PA 16802, USA}
}

\begin{abstract}
Pulsar winds shocked in the ambient
medium produce spectacular
nebulae
 observable from the
radio through $\gamma$-rays. The shape and the spectrum of a pulsar
wind nebula (PWN) depend on
the angular distribution, magnetization and energy spectrum of the
 wind streaming from the pulsar magnetosphere, as well as on the pulsar velocity and the properties
of the ambient medium. The advent of {\sl Chandra}, with its unprecedented  angular
resolution and high sensitivity, has allowed us not only to detect many
new PWNe, but also study their spatial and spectral structure and dynamics, which has
 significantly advanced our understanding of these objects.
 Here we  overview
recent observational
 results
 on PWNe, with emphasis on
{\sl Chandra} observations.
\end{abstract}


\maketitle


\section{Introduction}
It is generally accepted that all active pulsars lose their spin
energy and angular momentum
 via relativistic winds
comprised of relativistic particles and electromagnetic field.
Because the relativistic
bulk velocity of the wind
leaving  the pulsar magnetosphere
is obviously supersonic with respect to the ambient medium,
such a wind produces a {\em termination shock} (TS) at a distance $R_{\rm TS}$
from the pulsar where the bulk wind pressure,
$P_w\sim\edot/(4\pi c R_{\rm TS}^2)$,
is equal to the ambient pressure $P_{\rm amb}$.
The TS radius can be estimated as
\be
R_{\rm TS} \sim
\left(\edot/4\pi c P_{\rm amb}\right)^{1/2}
\sim
0.05\, \edot_{36}^{1/2} P_{\rm amb,-10}^{-1/2}\,\,\,\,{\rm pc},
\ee
where $\edot = 10^{36}\edot_{36}$ ergs s$^{-1}$ is the pulsar's spin-down power,
and $P_{\rm amb}=10^{-10} P_{\rm amb,-10}$ dyn cm$^{-2}$.
At the TS, the pulsar wind is being ``thermalized'',
and the downstream bulk flow speed becomes subrelativistic \cite{1984ApJ...283..710K}.
As the relativistic
particles of the shocked wind move in the magnetic field and ambient radiation
field,
they emit synchrotron
and inverse Compton (IC) radition, which we observe as a {\em pulsar wind
nebula}
(PWN).
Since the wind is a universal property of any active pulsar, we expect that
{\em all pulsars must be accompanied by PWNe}.
 Studying PWNe
tells us about the properties of pulsar
winds and  their parent pulsars,
the properties of the ambient
medium, and the mechanisms of wind-medium interaction.

The energies of the synchrotron and IC photons
span the range from the radio to TeV $\gamma$-rays.
The mere detection of a PWN in some energy band indicates
the emission mechanism and the electron energies involved. For instance,
 detecting a PWN in the X-ray band, where  the synchrotron emission dominates,
  implies that
the wind particles have
been accelerated up to $\gtrsim 100$ TeV energies, either at the TS or on the way to the TS
(note that particles
with such energies cannot leave the pulsar magnetosphere because of
radiative losses),
and that the same particles should produce IC emission in the TeV energy range.

Useful information on the pulsar wind and its interaction with the medium
is provided by the PWN morphology, which depends on the wind {\em outflow
geometry}
and the direction of {\em pulsar velocity}.
If we assume an {\em isotropic outflow} from a
very slowly moving pulsar, the shocked wind is confined between the TS
and {\em contact discontinuity} (CD) spheres,
while the shocked ambient medium fills in the space between the CD and
the spherical {\em forward shock} (FS).

If the pulsar moves with a supersonic velocity, $V_p \gg c_s$,
then the TS, CD and FS surfaces acquire a characteristic convex shapes ahead of
the moving pulsar but can exhibit  rather different shapes behind it.
In particular, the TS acquires a bullet-like shape, with
the distance
$R_{\rm TS,h}\sim (\edot/4\pi c P_{\rm ram})^{-1/2}\sim
0.04
\edot_{36}^{1/2}n^{-1/2}(V_p/100\,{\rm km\,s}^{-1})^{-1}$ pc,
between the bullet's head ({\em bowshock}) and the pulsar (cf.\ eq.\ 1),
 where $P_{\rm ram} = \rho V_p^2 = 1.67\times 10^{-10} n (V_p/100\, {\rm km\,s}^{-1})^2$
dyn cm$^{-2}$ is the ram pressure.
 The numerical simulations \cite{2003A&A...404..939V, 2005A&A...434..189B}
suggest that the shocked wind is channeled into the
{\em tail}
behind the TS bullet, confined by the nearly cylindrical CD surface,
where it flows with a mildly relativistic velocity.
 The shocked ISM matter
 is also stretched along the pulsar
trajectory,
and it can be seen
in spectral lines (e.g., H$_\alpha$) from the atoms
excited at the FS (e.g., \cite{2002ApJ...575..407C}).

\begin{figure}[h]
\label{p-pdot}
\hspace{-0.3in}
\includegraphics[height=3.3in,width=3.1in,angle=90]{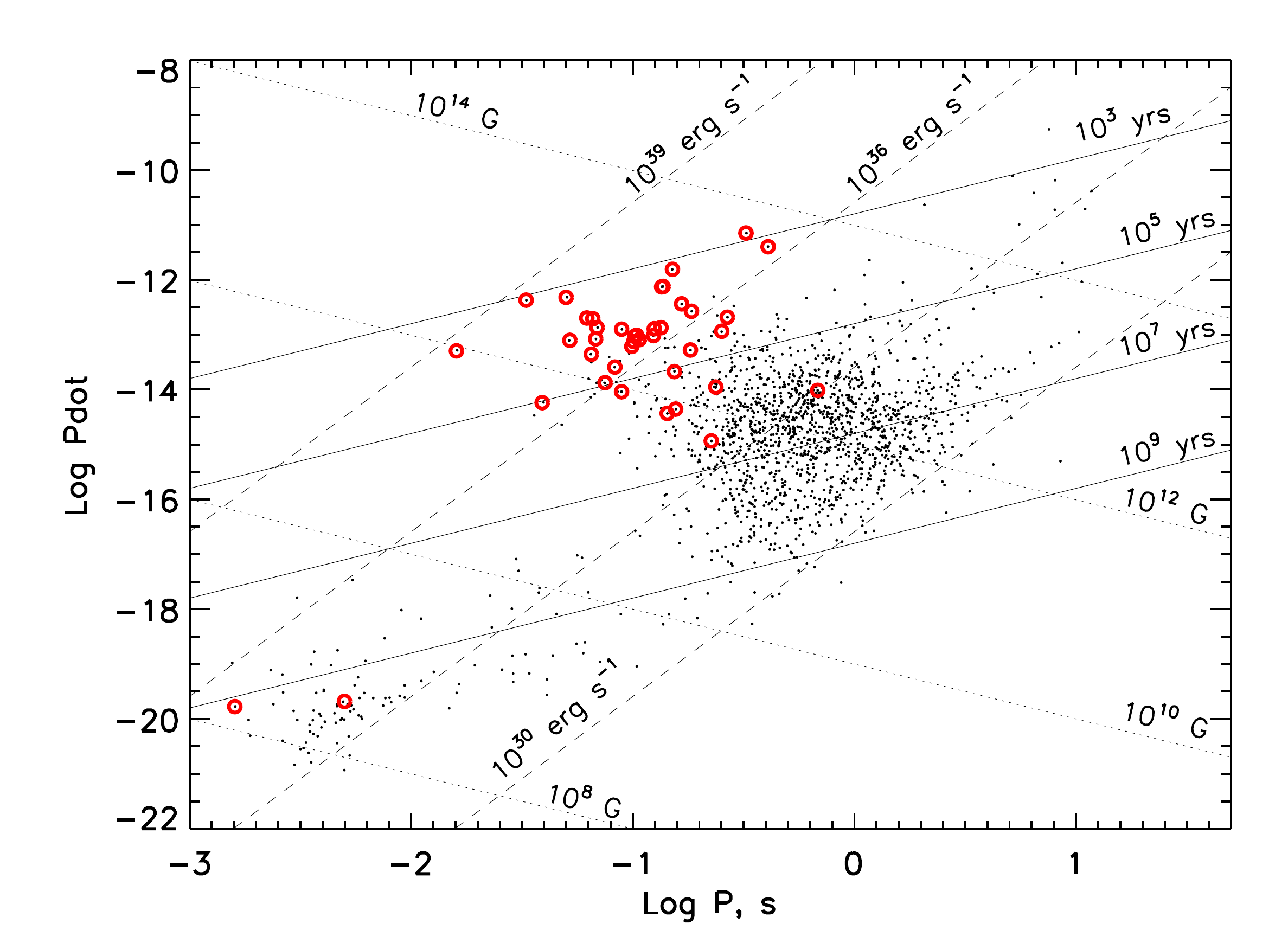}
\caption{$P$-$\dot{P}$ diagram for the pulsars in the ATNF catalog
\cite{2005AJ....129.1993M}.
Pulsars with known X-ray PWNe are marked by circles.
}
\end{figure}

  \begin{figure}
 \centering
\includegraphics[width=6.5in,angle=0]{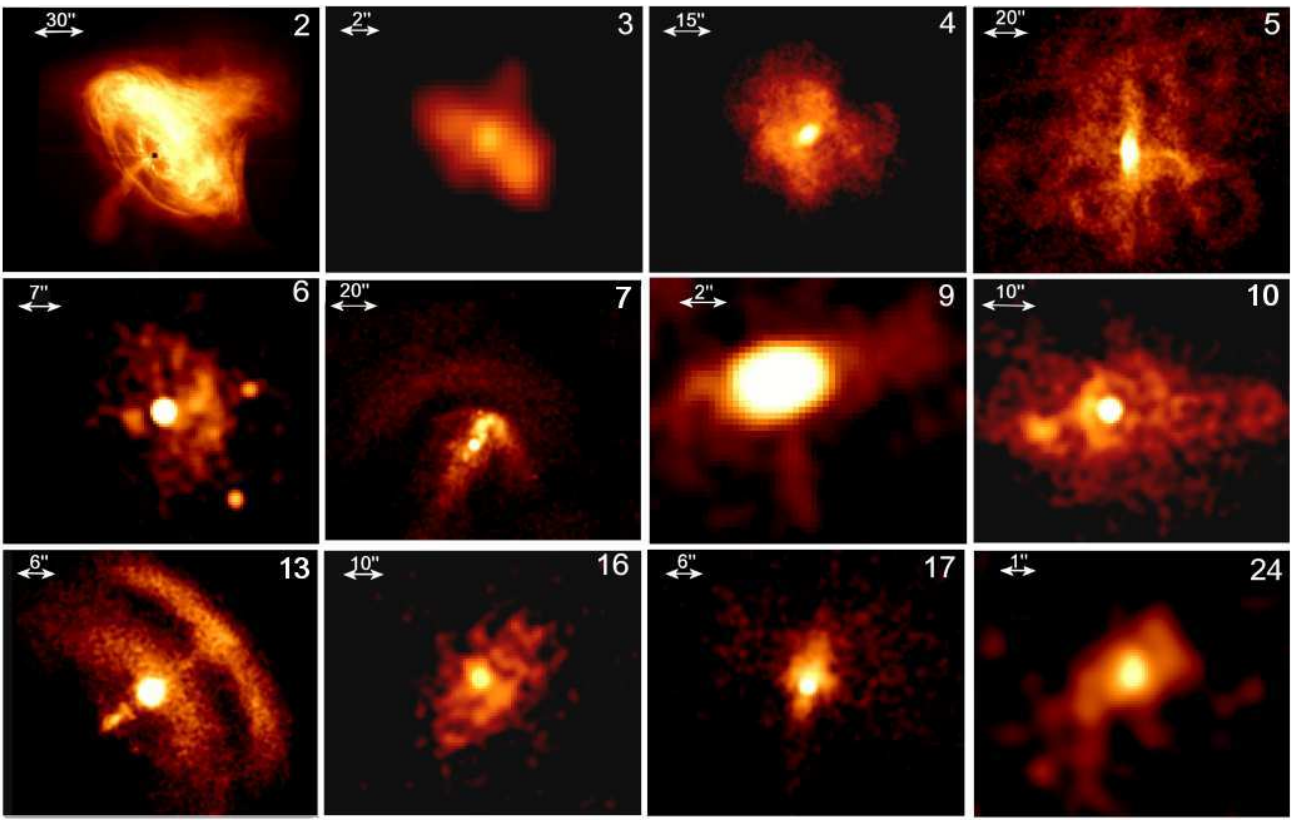}
\caption{X-ray images of PWNe with toroidal components. The image numbers
correspond to Tables 1--2. }
\end{figure}

We know from observations of young, subsonically moving pulsars
(such as the Crab pulsar), that the PWN is not spherical even in this
case, but it rather looks like a torus, sometimes with one or two
 jets along the pulsar's spin axis. This means that
the pre-shock wind is {\em not isotropic}, but it outflows preferentially in
the equatorial plane of the rotating pulsar. Models of such {\em torus-jet}
PWNe have been simulated in \cite{2004MNRAS.349..779K} and
 \cite{2006A&A...453..621D}  (see also Bucciantini, this proceedings).
We are not aware of relativistic MHD models for PWNe created by supersonically
moving pulsars with {\em anisotropic} outflows,
but we expect that the compact PWN morphology
near the pulsar may significantly
depend on the orientation of the spin axis with respect
to the direction of pulsar's motion, while the distant PWN tail
and the FS should not be
strongly affected by the wind anisotropy.

As the PWN appearance critically depends on the pulsar's Mach number,
${\mathcal M}=V_p/c_s$, and $c_s$ in a hot SNR interior is much
higher than that in the normal ISM, we expect that young, powerful pulsars,
which have not left their host SNRs,
generate torus-jet PWNe, while older pulsars
would show bowshock-tail
PWNe \cite{2003A&A...397..913V}.

Only a handful of PWNe had been detected before the launch of \chan,
mostly in radio and H$_\alpha$ observations \cite{2002ApJ...575..407C},
while X-ray observations of PWNe had been hindered by low angular resolution
of X-ray instruments.
\chan, with its unprecedented $\approx 0.5''$
resolution, has allowed us to reveal the
fine structure of the previously known X-ray PWNe, discover many new
PWNe, and separate the pulsar and PWN emission in many cases.
Some interesting results on X-ray PWNe have also been obtained with
{\sl XMM-Newton}, which lacks the high angular resolution
of \chan\ but is more sensitive
and has a larger field of view.
Many exciting results from the \chan\ and \xmm\ observations of PWNe
have been reported in numerous publications, including two reviews
\cite{2006csxs.book..279K, 2006ARA&A..44...17G}.
Here, we present an up-to-date overview of
X-ray observations
of PWNe,
including a gallery of spectacular PWN images and a current catalog of these objects,
and discuss correlations between various PWN properties as well as
the pulsar-PWN correlations.

\section{
Chandra images}

Fifty four PWNe detected with \chan\ are  listed in Tables 1--3.
Forty of these PWNe are powered by known pulsars
(marked by circles in Fig.\ 1), while the remaining  fourteen have not been
associated with  pulsars so far, but their properties
strongly suggest a PWN origin.

The gallery of  X-ray images shown in Figures 2--4
 demonstrates
the  amazing diversity of PWN shapes and
morphologies\footnote{See \url{http://www.astro.psu.edu/users/green/pwne/pwne.html} for high-resolution
color images.}.
Yet, among the bright, well-resolved   PWNe
two morphological types
 can be
distinguished:
 {\em torus-jet PWNe}, which
show an elliptical (toroidal) structure around
the pulsar and sometimes one or two jets
along the torus axis (Fig.\ 2), and {\em bowshock-tail PWNe}, whose appearance is dominated by
a cometary structure, with the pulsar close to the ``comet head'' (Fig.\ 3).

An archetypical example of a torus-jet PWN is
the famous {\em Crab PWN} (\#2; hereafter
we refer to a particular PWN using the numbering scheme
introduced
in Tables 1--3). Its
\chan\ image \cite{2000ApJ...536L..81W}
shows an axisymmetric PWN
with a tilted {\em inner ring}, associated with the TS
in the equatorial pulsar wind,
a {\em torus} (with numerous thin wisps in its inner part),
associated with the shocked
pulsar wind, and {\em two jets} emanating along the pulsar spin
axis. Remarkably, the jets are stretched approximately along
the direction of the pulsar's proper motion,
implying that the pulsar
got a ``kick'' along its spin
axis in the process of formation (e.g., \cite{1998Natur.393..139S}).
Multiple \chan\ observations have
 revealed the remarkable dynamics of the Crab
  \cite{2002ApJ...577L..49H}\footnote{\url{http://chandra.harvard.edu/photo/2002/0052/movies.html}}.
  In particular,
 these observations have shown that the wisps are being created
at the inner ring and propagate outwards with a speed of $\sim 0.5 c$,
while the motions further away in the torus seem to slow down with distance from the pulsar.
Also, the jets show variability on a much longer time scale of months
 \cite{2004IAUS..218..181M}.

  \begin{figure}[ht!]
 \centering
\includegraphics[width=6.5in,angle=0]{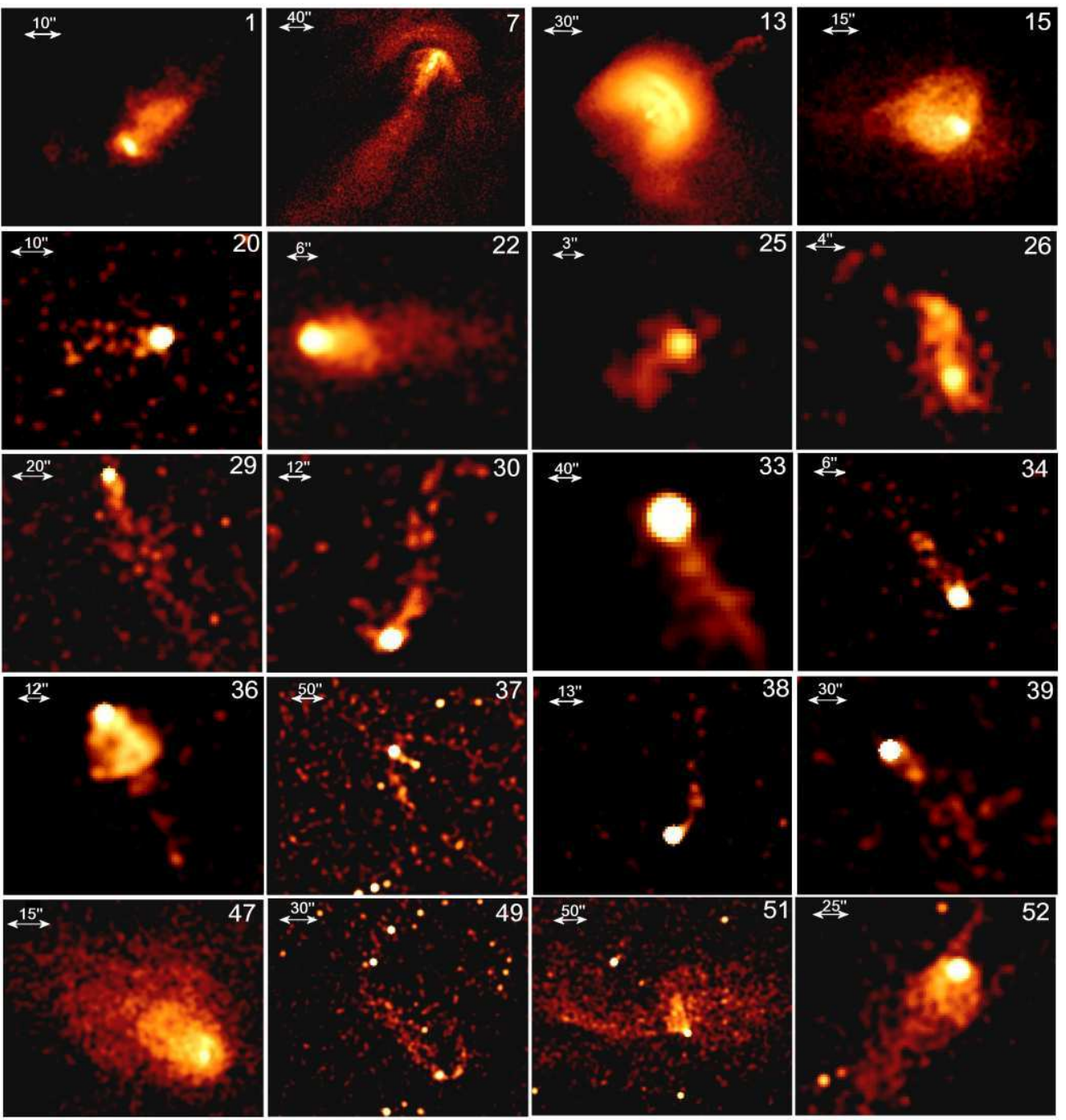}
\caption{X-ray images of PWNe whose shapes are affected by the pulsar motion.
The numbers correspond to Tables 1--3.}
\end{figure}

In Figure 2 we see more examples of PWNe with (presumably) toroidal
components and, in most cases,
jets, with one jet often being much brighter then the other,
 possibly due to the Doppler boosting effect (the approaching jet looks brighter
than the receding one). However, appearrance of some of the PWNe is noticeably different from the Crab.
For example, in addition to the compact elongated core
(possibly a  torus projected edge-on),
the 3C\,58 SNR (\#5; \cite{2004ApJ...616..403S}),
shows multiple
loop-like filaments,
suggesting a
complex structure of the magnetic field
and instabilities in the shocked wind.
An unusual morphology, with two arcs (possibly part of two rings above
and below the equatorial plane \cite{2001ApJ...556..380H}), bright inner
jets, and much fainter, strongly variable outer jets
\cite{2003ApJ...591.1157P},
is seen in the
Vela PWN (\#13; we will discuss it below in more detail).
Another example is the
``Jellyfish'' (\#7), powered by the young PSR B1508--58,
which shows
a bizarre structure with two ``semi-arcs'' (one of which is perhaps a curved jet)
and a very long tail
(a jet?)
southeast of the pulsar
 \cite{2002ApJ...569..878G}.
Interestingly, there are pairs of young PWNe in SNR (e.g., \#4 and \#5, \#9 and \#10)
with quite different appearance despite
very similar powers and ages of their parent pulsars.
This dissimilarity can possibly be attributed
to different inclinations of the pulsar's spin and magnetic axes
(hence different PWN projections onto the sky plane).
Furthermore, in several cases we see a mixture of the toroidal and cometary
morphologies. For instance, the PWN generated by the most powerful
PSR J0537--6910 (\#1, shown in Fig.\ 3)
exhibits, in addition to a compact torus-like component, a huge,
$\sim$4 pc long,
cometary structure,
likely a bubble of shocked relativistic wind behind the high-speed pulsar
\cite{2001ApJ...559..275W, 2006ApJ...651..237C}.
Also, the overall appearance of the Vela and Jellyfish PWNe
is obviously affected by
the pulsar motion (that is why we show
these objects in both Fig.\ 2 and Fig.\ 3, at different scales).

  \begin{figure}[ht!]
 \centering
\includegraphics[width=6.5in,angle=0]{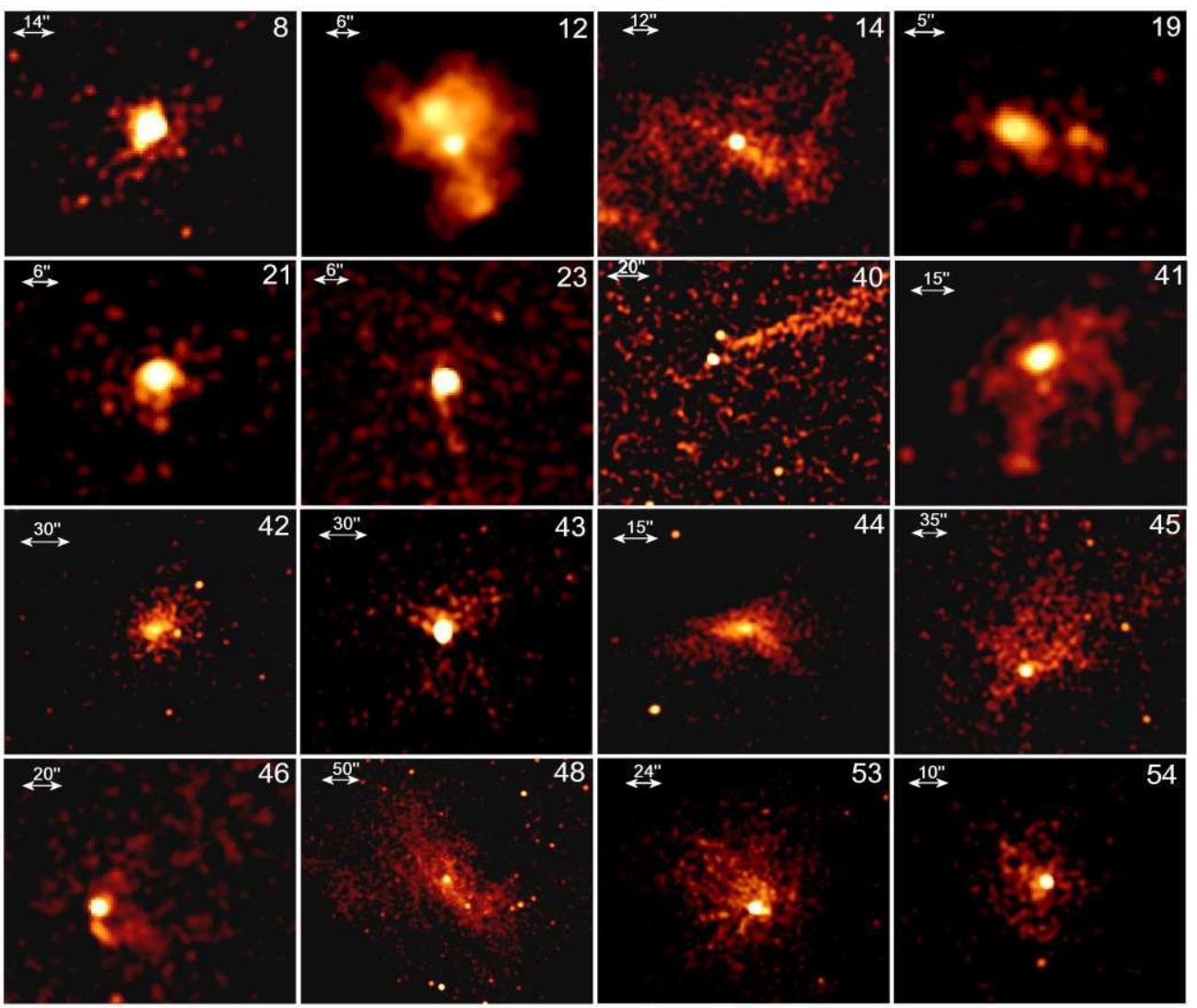}
\caption{X-ray images of PWNe with uncertain morphologies. The numbers correspond
to Tables 1--3.}
\end{figure}

The cometary structure of the PWNe shown in Figure 3 implies that they are
shaped by the pulsar's motion in the ambient medium. For most of these PWNe,
we can assume that the pulsar moves supersonically, so that the
brightest part of the
PWN is associated with the shocked wind
just outside the TS ``bullet'', with the pulsar close to the bullet's head,
while the
tail
behind the bullet represents the shocked wind
confined by the nearly cylindrical CD surface (see Introduction).
A typical example of
such a bowshock-tail PWN is the ``Mouse'' (\#22) \cite{2004ApJ...616..383G}),
which looks even more spectacular in the radio range \cite{2005AdSpR..35.1129Y}.
Another outstanding
example
J1509--5058 (\#29), with its extremely long,
$\gtrsim$6 pc, tail (we will discuss this and similar objects below).

As expected (see Introduction), most of the toroidal PWNe are powered by young pulsars,
and the oldest PWNe without an identifiable SNR exhibit bowshock-tail morphology.
However, there
is no  strict correlation with the age and power, nor with the presence of an SNR.
Not only we see cometary structures in some young PWNe in SNRs
(e.g., \#\# 1, 7, 13, 47), but also there are  a few older PWNe, not associated
with SNRs, which look like typical torus-jet PWNe
(e.g., \#16; \cite{2004ApJ...612..389H}).

We should also note that the morphologies of some of the cometary PWNe
are very different from those expected from the
current MHD models.
For example, the  ``Mushroom'' PWN around
PSR B0355+54 (\#36; \cite{2006ApJ...647.1300M}) consists of a broad, bright
``cap''
 and a narrow, faint ``stem''.
Another unusual example is the Geminga PWN (\#37), which shows a shell-like structure
with a bow head and a cylindrical body seen up to $\sim$0.2 pc behind
the pulsar, and a short (0.05 pc), narrow tail (a jet?)
along the symmetry axis of the shell \cite{2003Sci...301.1345C, 2006ApJ...643.1146P}.
Such a picture
suggests that the Geminga's wind is essentially anisotropic, possibly
concentrated around the equatorial plane perpendicular to the pulsar's velocity.
Very peculiar is the PWN of the recycled pulsar J2124--3358 (\#38),
which shows a curved X-ray tail within an asymmetric H$_\alpha$
bowshock,
misaligned with each other and the direction of pulsar's proper motion
\cite{2002ApJ...580L.137G,
2005AAS...20718313C, 2006AnA...448L..13H}. Such a structure might imply
nonuniformities in the ambient ISM, in addition to anisotropy of the pulsar wind.

 \begin{figure}[h]
 \centering
\hspace{-0.2cm}
\includegraphics[width=.44\textwidth,angle=0]{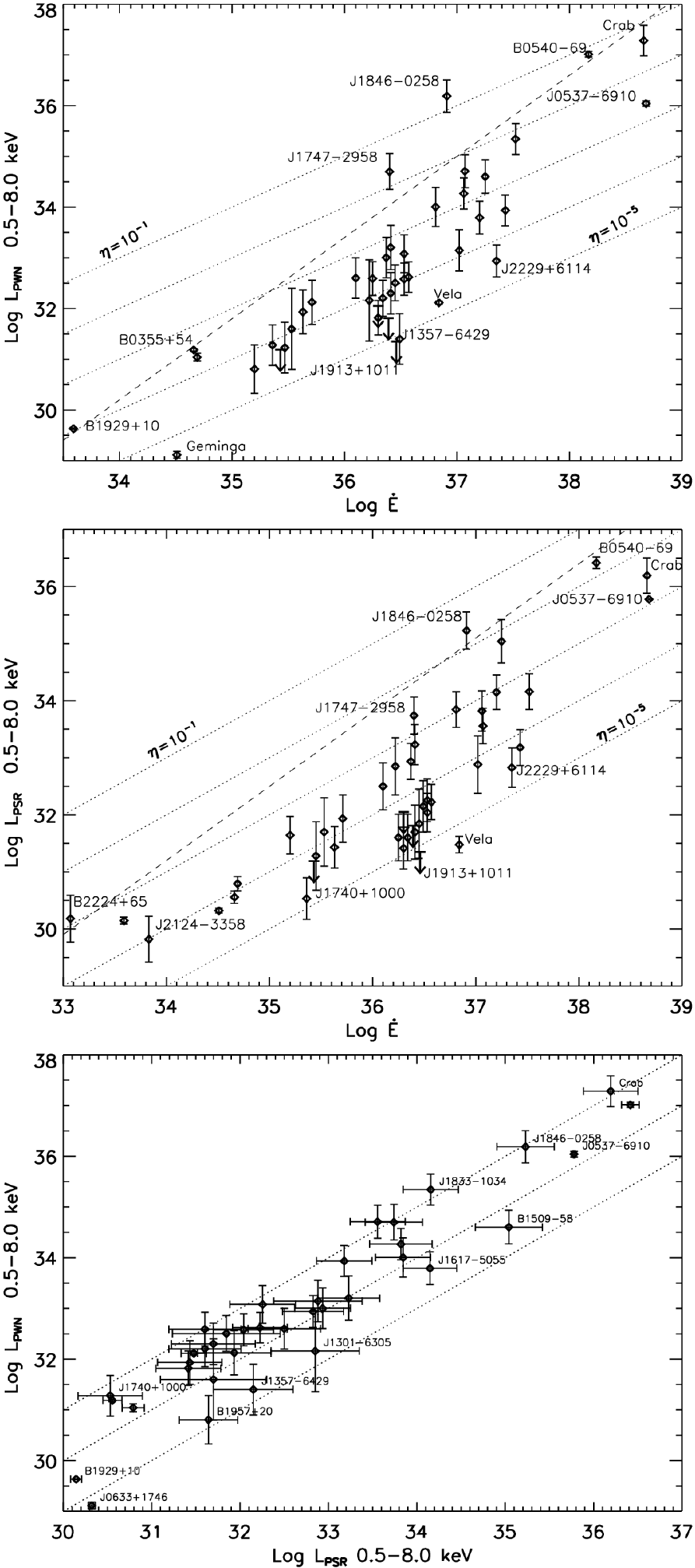}
\caption{
Dependences of the PWN luminosity and nothermal component of the pulsar luminosity
on the pulsar spin-down power ({\em top} and {\em middle} panels, respectively).
The dotted lines correspond to constant X-ray efficiencies, $\eta=L/\edot$.
The dashed lines, $\log \lpwn=1.6 \log \edot -24.2$ and $\log \lpsr=1.3 \log \edot  -13.0$,
 show approximate upper bounds
for majority of objects.
The lower panel demonstrates the correlation between the pulsar and PWN luminosities.}
\end{figure}

\begin{table}
 \setlength{\tabcolsep}{0.05in}
 \begin{tabular}{rlccrccccccc}
\hline
 \tablehead{1}{c}{b}{   \# \\  }  &
  \tablehead{1}{c}{b}{  ${\rm PSR}$\tablenote{The superscript $^{\rm T}$ marks pulsars with a thermal component in the {\sl Chandra} band.}  \\  } &
   \tablehead{1}{c}{b}{  ${\rm SNR}$ \\  } &
    \tablehead{1}{c}{b}{  ${\rm PWN}$ \\  } &
      \tablehead{1}{c}{b}{  $\log\dot{E}$ \\  } &
   \tablehead{1}{c}{b}{  $ P$ \\  ${\rm ms}$ } &
      \tablehead{1}{c}{b}{  $\log\tau$ \\  } &
     \tablehead{1}{c}{b}{  $\log B_{s}$\tablenote{Logarithm of magnetic field at the neutron star surface. } \\  } &
      \tablehead{1}{c}{b}{  $\log B_{\rm LC}$\tablenote{Logarithm of magnetic field at the light cylinder. } \\  } &
   \tablehead{1}{c}{b}{  $d$\tablenote{Our best guess for the pulsar distance, used to scale the distance-dependent parameters in Table \ref{tab:prop2}. The subscript $^{\rm p}$ marks the distances determined by parallax measurements.} \\  ${\rm kpc}$ } &
    \tablehead{1}{c}{b}{  ${\rm Rad./H}_{\alpha}{\rm TeV}$\tablenote{Is the PWN detected in radio, H$_{\alpha}$, and TeV $\gamma$-rays? P = `possibly'. } \\  }
      \\
\hline
1   &   J0537--6910           &                N157B     &   N157B            &   38.68   &   16        &   3.70   &   12.97 &   6.31 &     50~~~~~      &  Y/N/N     \\
2   &   B0531+21              &  G184.6--5.8/Crab        & Crab               &   38.66   &   33        &   3.09   &   12.58 &   5.99 &    2.0              &  Y/N/Y      \\
3   &   B0540--69             &                N158A     &    N158A           &   38.17   &   50        &   3.22   &   12.70 &   5.56 &     50~~~~~       &  Y/N/N     \\
4   &   J1833--1034           &  G21.5--0.9             &   G21.50--0.89      &   37.52   &   62        &   3.69   &   12.55 &   5.15 &    4.7              &  Y/N/Y      \\
5   &   J0205+6449$^{\rm T}$  &  G130.7+3.1/3C\,58       &   3C\,58           &   37.43   &   66        &   3.73   &   12.56 &   5.07 &    3.2              &  Y/N/N      \\
6   &   J2229+6114$^{\rm T}$  &  G106.3+2.7              &  G106.65+2.96      &   37.35   &   52        &   4.02   &   12.31 &   5.14 &    3~~~                &  Y/N/N      \\
7   &   B1509--58             &  G320.4--1.2/MSH\,15--52 &   Jellyfish        &   37.25   &  151        &   3.19   &   13.19 &   4.62 &     5~~~               &  P/N/Y      \\
8   &   J1617--5055           &  ...                     &  G332.50--0.28     &   37.20   &   69        &   3.91   &   12.49 &   4.94 &     6.5             &  N/N/P      \\
9   &   J1124--5916           &  G292.0+1.8/MSH\,11--54  &   G292.04+1.75     &   37.07   &  135        &   3.46   &   13.01 &   4.59 &     6~~~               &  Y/N/N      \\
10  &   J1930+1852            &  G54.1+0.3               &   G54.10+0.27      &   37.06   &  137        &   3.46   &   13.01 &   4.57 &    5~~~             &  Y/N/N      \\
11  &   J1420--6048            &  ...                    &  G313.54+0.23      &   37.02   &   68        &   4.11   &   12.38 &   4.85 &    5.6              &  P/N/Y      \\
12  &   J1846--0258           &  G29.7--0.3/Kes\,75      &  Kes\,75           &   36.91   &  324        &   2.86   &   13.69 &   4.12 &     19~~~~~	             &  Y/N/Y      \\
13  &   B0833--45$^{\rm T}$   &  G263.9--3.3/Vela        & Vela               &   36.84   &   89        &   4.05   &   12.53 &   4.65 &   ~~~~$0.29^{\rm p}$    &  Y/N/Y      \\
14  &   J1811--1925           &  G11.2--0.3              &   G11.18--0.35     &   36.81   &   65        &   4.37   &   12.23 &   4.77 &    5~~~            &  Y/N/N      \\
15  &   B1951+32              &  G69.0+2.7/CTB\,80       &  G68.77+2.82       &   36.57   &   39        &   5.03   &   11.69 &   4.87 &    2.5              &  Y/Y/N      \\
16  &   J2021+3651$^{\rm T}$  &  ...                     &   G75.23+0.12      &   36.53   &  104        &   4.23   &   12.50 &   4.43 &    4~~~            &  N/N/N     \\
17  &   B1706--44$^{\rm T}$   &  G343.1--2.3             &   G343.10-2.69     &   36.53   &  102        &   4.24   &   12.49 &   4.43 &     2~~~            &  Y/N/N     \\
18  &   J1357--6429$^{\rm T}$ &  ...                     &  G309.92--2.51     &   36.49   &   166       &   3.86   &   12.90 &   4.20 &    2.5              &  P/N/N      \\
19  &   B1823--13$^{\rm T}$   &  ...                     &  G18.00--0.69      &   36.45   &   101       &   4.33   &   12.45 &   4.40 &   4~~~            &  N/N/Y      \\
20  &   B1757--24             &  ...                     &            Duck    &   36.41   &  125        &   4.19   &   12.61 &   4.29 &        5~~~  &  Y/N/N      \\
21  &   J1016--5857$^{\rm T}$ &  G284.3-1.8              &   G284.08--1.88    &   36.41   &   107       &   4.32   &   12.47 &   4.35 &        3            &  N/N/N      \\
22  &   J1747--2958           &  ...                     &             Mouse  &   36.40   &   99        &   4.41   &   12.40 &   4.38 &        5~~~   &  Y/N/N      \\
23  &   J1119--6127$^{\rm T}$ &  G292.2-0.5              &    G292.15--0.54   &   36.37   &  408        &   3.21   &   13.61 &   3.75 &   8.4             &  N/N/N      \\
24  &   B1800--21$^{\rm T}$   &  ...                     &  G8.40+0.15        &   36.34   &  134        &   4.20   &   12.63 &   4.22 &    4~~~            &  N/N/P      \\
25  &   B1046--58             &  ...                     &  G287.42+0.58      &   36.30   &  124        &   4.31   &   12.54 &   4.24 &   3~~~            &  N/N/N     \\
26  &   J1809--1917$^{\rm T}$ &  ...                     &  G11.09+0.08       &   36.25   &   83        &   4.71   &   12.17 &   4.39 &    3.5              &  P/N/P       \\
27  &   J1301--6305           &   ...                    &  G304.10--0.24     &   36.22   &  184        &   4.04   &   12.85 &   4.02 &    7~~~                &  N/N/Y       \\
28  &   J1718--3825            &  ...                    &  G348.95--0.43    &   36.10   &   75        &   4.95   &   12.00 &   4.35 &    4~~~             &  N/N/Y       \\
29  &   J1509--5850            &  ...                    &  G319.97--0.62     &   35.71   &   89        &   5.19   &   11.96 &   4.09 &    4~~~           &  P/N/N       \\
30  &   B1853+01              &  G34.7--0.4/W44          &   G34.56-0.50      &   35.63   &   267       &   4.31   &   12.88 &   3.57 &   3~~~        &  Y/N/N       \\
31  &   J1702--4128           &  ...                     &  G344.74+0.12      &   35.53   &   182       &   4.74   &   12.49 &   3.69 &    5~~~        &  N/N/Y      \\
32  &   J0729--1448            &  ...                    &  G230.39--1.42     &   35.45   &   252       &   4.54   &   12.73 &   3.50 &     4~~~      &  N/N/N       \\
33  &   J1740+1000$^{\rm T}$  &  ...                     &  G34.01+20.27      &   35.36   &   154       &   5.06   &   12.27 &   3.67 &    1.4              &  N/N/N       \\
34  &   B1957+20              &  ...                     &  G59.20--4.70      &   35.20   &   1.6       &   9.18   &   ~\,8.22&  5.57 &    2.5              &  N/Y/N      \\
35  &   J0538+2817$^{\rm T}$  &  G180.0--1.7/S147        &  G179.72--1.69     &   34.69   &   143       &   5.79   &   11.86 &   3.37 &   ~~~~$1.47^{\rm p}$    &  N/N/N      \\
36  &   B0355+54$^{\rm T}$    &  ...                     &       Mushroom     &   34.66   &   156       &   5.75   &   11.92 &   3.31 &   ~~~~$1.04^{\rm p}$    &  N/N/N       \\
37  &   J0633+1746$^{\rm T}$  &  ...                     &      Geminga       &   34.51   &   237       &   5.53   &   12.21 &   3.06 &    ~~~~$0.25^{\rm p}$   &  N/N/N       \\
38  &   J2124--3358            &  ...                    &  G10.93--45.44     &   33.83   &   5         &   9.58   &   ~\,8.51&  4.40 &  ~~~0.25           &  N/Y/N       \\
39  &   B1929+10$^{\rm T}$     &  ...                    &  G47.38--3.88      &   33.59   &   226       &   6.49   &   11.71 &   2.62 &  ~~~~$0.36^{\rm p}$   &  P/N/N     \\
40  &   B2224+65              &  ...                     &          Guitar    &   33.07   &   683       &   6.05   &   12.41 &   1.88 &     1~~~       &  N/Y/N      \\\hline
\end{tabular}
\caption{Pulsars with X-ray PWNe.}
\label{tab:prop1}
\end{table}
\normalsize

\begin{table}
 \setlength{\tabcolsep}{0.1in}
\begin{tabular}{rlcccclllc}
\hline
 \tablehead{1}{c}{b}{   \# \\  }  &
  \tablehead{1}{c}{b}{   $n_{\rm H,22}$\tablenote{Hydrogen column density (in units of $10^{22}$ cm$^{-2}$)
obtained from spectral fits to the PWN spectra or
 estimated  from the pulsar's dispersion measure assuming 10\% ISM ionization (in square brackets for the latter case).} \\  }  &
   \tablehead{1}{c}{b}{   $\log L_{\rm pwn}$\tablenote{Logarithm of
 PWN luminosity in the 0.5--8 keV band, in units of  ergs
s$^{-1}$.  For
bright PWNe
(e.g., \#\# 2, 5, 13),  we quote
the luminosity of the PWN ``core''
restricted to the torus/arcs
regions.
For the PWNe with extended tails
 (\#\# 29, 30, 33, 35, 36, 37, 39)
we quote only the luminosity of the bright ``bullet'' component,
 while the tail luminosities are listed in Table \ref{tab:tails}).
For \#\# 27, 31 and 32, faint extended emission is seen around the pulsar but its luminosity is
very uncertain; we use $\pm 0.50$ as a conservative estimate. No compact PWN is resolved for
\#38 and \#40, but tails are possibly seen (see text and Table \ref{tab:tails}). } \\  }  &
   \tablehead{1}{c}{b}{   $\Gamma_{\rm pwn}$  \\  }  &
   \tablehead{1}{c}{b}{   $\log L_{\rm psr}^{\rm nonth}$~\tablenote{Logarithm of nonthermal pulsar luminosity
in the 0.5--8 keV band, in units of
ergs s$^{-1}$.
In the cases when the spectrum
is fitted with
the blackbody+powerlaw model (for the pulsars marked with $^{\rm T}$ in Table 1), it is the luminosity of
the power-law component only. } \\  }  &
    \tablehead{1}{c}{b}{   $\Gamma_{\rm psr}$  \\  }  &
      \tablehead{1}{c}{b}{   $l_{X}$\tablenote{
Characteristic size of the PWN ``core''
in which the PWN X-ray properties listed in this table were measured.}  \\  ${\rm pc}$ }  &
  \tablehead{1}{c}{b}{   $R_{\rm TS}$\tablenote{Estimated TS stand-off
distance.
}  \\ ${\rm pc}$ }  &
 \tablehead{1}{c}{b}{   $P_{\rm amb,-9}$\tablenote{Estimated ambient pressure, in units of $10^{-9}$ dyn cm$^{-2}$,
assuming an isotropic wind (see eq.\ 1).}  \\ } &
\tablehead{1}{c}{b}{ ${\rm Refs.}$\tablenote{
The PWN/PSR X-ray properties listed here were measured by ourselves
(except for \#2, \#7, and \#28), but
we cite recent relevant papers when available.}\\}
  \\
\hline
1       &     0.5   &    $36.04  \pm   0.01$    &     $2.20\pm0.05$       &   $35.78  \pm  0.01$   &      $1.8\pm0.1$            &  1.4  &   $ 0.19  $ &   $   3.7  $    &  \cite{2006ApJ...651..237C}    \\
2       &     0.32  &    $37.28  \pm   0.01$    &     $2.12\pm0.01$       &   $36.19  \pm  0.01$   &      $1.63\pm0.09$          &  1.2  &   $ 0.14  $ &   $
                        6.6$    &  \cite{2004ApJ...609..186M, 2001AnA...365L.212W}    \\
3       &     0.46  &    $37.01  \pm   0.01$    &     $1.85\pm0.10$       &   $36.41  \pm  0.10$   &      $2.05\pm0.08$          &  1.4  &   $ 0.24  $ &   $   0.72 $    &  \cite{2001ApJ...546.1159K}    \\
4       &     2.3   &    $35.34  \pm   0.01$    &     $1.89\pm0.02$       &   $34.16  \pm  0.01$   &      $1.51\pm0.07$          &  1.0  &   $\negsp <0.05$  &   $\negsp >3.7 $    &  \cite{2001ApJ...561..308S}   \\
5       &     0.43  &    $33.94  \pm   0.01$    &     $2.02\pm0.01$       &   $33.18  \pm  0.01$   &      $1.7\pm0.04$           &  1.2 &   $  0.054$ &   $   2.6  $    &  \cite{2004ApJ...616..403S}    \\
6       &     0.5   &    $32.94  \pm   0.01$    &     $1.3\pm0.1$         &   $32.83  \pm  0.05$   &      $1.5\pm0.1$            &  0.4  &   $ 0.07  $ &   $  1.3   $    &  \cite{2001ApJ...552L.125H}    \\
7       &     0.8   &    $34.60  \pm   0.03$    &     $1.65\pm0.05$       &   $35.04  \pm  0.08$   &      $1.2\pm0.1$            &  4.5  &   $  0.4  $ &   $   0.03 $    &  \cite{2002ApJ...569..878G}  \\
8       &     3.5   &    $33.79  \pm   0.02$    &     $1.2\pm0.2$         &   $34.15  \pm  0.01$   &      $1.15\pm0.10$          &  0.6  &   $  0.05 $ &   $   1.8  $    & ...     \\
9       &     0.37  &    $34.71  \pm   0.03$    &     $1.7\pm0.5$         &   $33.56  \pm  0.01$   &      $1.68\pm0.05$          &  0.2  &   $\negsp <0.15 $ &   $\negsp >0.15  $    & \cite{2001ApJ...559L.153H} \\
10      &     1.9   &    $34.27  \pm   0.01$    &     $1.99\pm0.03$       &   $33.82  \pm  0.05$   &      $1.34\pm0.07$          &  1.2  &   $  0.14 $ &   $   0.17 $    & \cite{2002ApJ...568L..49L}  \\
11      &     5.4   &    $33.15  \pm   0.11$    &     $0.5\pm 1.2$
                                                                       &   $32.88  \pm  0.20$   &      $1.0\pm0.5$            &  0.4  &   $\negsp <0.1  $ &   $\negsp >  0.3$    &  \cite{2005ApJ...627..904N} \\
12      &     4.0   &    $36.19  \pm   0.02$    &     $2.03\pm0.02$       &   $35.23  \pm  0.03$   &      $1.30\pm0.06$          &  2.8  &   $\negsp <0.09 $ &   $\negsp >
                                      0.3$    & \cite{2003ApJ...582..783H}  \\
13      &     0.02  &    $32.11  \pm   0.03$    &     $1.4\pm0.1$         &   $31.48  \pm  0.15$   &      $2.0\pm0.3$            &  0.1  &   $ 0.016 $ &   $  8.0   $    &    \cite{2001ApJ...554L.189P, 2001ApJ...556..380H} \\
14      &     3.1   &    $34.00  \pm   0.09$    &     $1.5\pm0.2$         &   $33.84  \pm  0.01$   &      $1.4\pm0.1$            &  1.0  &   $\negsp <0.05 $ &   $\negsp >
                                      0.7$    & \cite{2003ApJ...588..992R}  \\
15      &     0.34  &    $32.62  \pm   0.01$    &     $1.76\pm0.03$       &   $32.22  \pm  0.01$   &      $1.70\pm0.03$          &  0.4 &   $\negsp <0.01 $ &   $\negsp >10   $    &  \cite{2004ApJ...610L..33M, 2005ApJ...628..931L}  \\
16      &     0.7   &    $33.08  \pm   0.07$    &     $1.7\pm0.3$         &   $32.25  \pm  0.07$   &      $1.0^{+0.6}_{-0.3}$    &  0.8  &   $ 0.068 $ &   $  0.22  $    & \cite{2004ApJ...612..389H} \\
17      &     0.5   &    $32.58  \pm   0.02$    &     $1.8\pm0.1$         &   $32.04  \pm  0.04$   &      $1.7\pm0.2$            &  0.2  &   $ 0.012 $ &   $   7.4  $    & \cite{2005ApJ...631..480R}  \\
18      &    0.23   &    $31.40   \pm  0.20$    &      ...                &   $32.15   \pm 0.15$   &      $1.3\pm0.2$            &  0.03 &   $\negsp <0.001$ &   $\negsp >9   $    & \cite{2007ApJ...665L.143Z}   \\
19      &     1.0   &    $32.50  \pm   0.05$    &     $1.3\pm0.4$         &   $31.84  \pm  0.31$   &      $1.9\pm0.7$            &  0.2  &   $\negsp <0.04 $ &   $\negsp >
                                      0.6$    &  \cite{2003ApJ...588..441G, 2007arXiv0707.3529P}  \\
20      &     4.4   &    $33.20  \pm   0.14$    &     $2.5\pm0.3$         &   $33.23  \pm  0.05$   &      $1.9\pm0.3$            &  0.5  &   $  0.05 $ &   $   0.3  $    &  \cite{2001ApJ...562L.163K}   \\
21      &  $\!\!\!$[1.2] &    $32.30  \pm   0.11$    &     $1.5\pm0.2$         &   $31.70  \pm  0.17$   &      $1.5\pm 0.4$           &  0.1  & $\negsp <0.03 $ &   $\negsp >
                                         0.9$    &  \cite{2004ApJ...616.1118C}   \\
22      &     3.0   &    $34.70  \pm   0.05$    &     $2.0\pm0.2$         &   $33.74  \pm  0.02$   &      $1.6\pm0.1$            &  0.5  &   $  0.02 $ &   $   1.7  $    &  \cite{2004ApJ...616..383G}   \\
23      &     1.6   &    $33.00  \pm   0.10$    &
                                                  $1.5\pm0.3$        &   $32.93  \pm  0.02$   &      $1.5_{-0.2}^{+0.3}$  &  0.5  &   $\negsp <0.1  $ &   $\negsp >0.06 $    &  \cite{2003ApJ...591L.143G}   \\
24      &     1.4   &    $32.20  \pm   0.05$    &     $1.6\pm0.3$         &   $31.60  \pm  0.11$   &      $1.4\pm0.6$            &  0.2  &   $  0.02 $ &   $   0.7  $    &  \cite{2007ApJ...660.1413K}  \\
25      &  $\!\!\!$[0.4] &    $31.82  \pm   0.04$    &     $1.0\pm0.2$         &   $31.41  \pm  0.07$   &      $1.5\pm0.3$            &  0.2  &   $ 0.024 $ &   $   1.0  $    &  \cite{2006ApJ...652..569G}   \\
26      &     0.7   &    $32.59  \pm   0.03$    &     $1.4\pm0.1$         &   $31.60  \pm  0.11$   &      $1.2\pm0.6$            &  0.2 &   $\negsp <0.05 $ &   $\negsp >
                                      0.2$    &  \cite{2007ApJ...670..655K}    \\
27      &  $\!\!\!$[1.1] &    $32.16   \pm  0.50$    &     ...                 &   $32.85   \pm 0.20$   &      ...                    &  2.0  &    ...   &    ...      &  ...  \\
28      &     0.7   &    $32.60   \pm  0.10$    &     $1.9\pm0.2$         &   $32.50   \pm 0.11$   &      $1.4\pm0.2$            &  2.0    &    ...    &     ...       &  \cite{2007arXiv0710.0367H}    \\
29      &     2.1   &    $32.12  \pm   0.13$    &     $1.8\pm0.3$         &   $31.93  \pm  0.12$   &      $2.0\pm0.3$            &  0.4  &   $\negsp <0.02 $ &   $\negsp >0.36 $    &  \cite{2007AandA...470..965H}    \\
30      &     2.0   &    $31.93  \pm   0.13$    &     $2.1\pm1.0$         &   $31.43  \pm  0.06$   &      $1.4\pm0.2$            &  0.4  &   $\negsp <0.04 $ &   $\negsp >0.07 $    &  \cite{2002ApJ...579..404P}    \\
31      & $\!\!\!$[1.1] &    $31.60   \pm   0.50$   &      ...                &   $31.70   \pm  0.30 $     &         ...                &  0.2  &   $ ...   $ &   $  ...   $    &  ...  \\
32      & $\!\!\!$[0.3] &    $31.20\pm 0.50$       &      ...                &   $31.28   \pm  0.30$      &      ...                   &  0.05 &   $ ...   $ &   $   ...  $    &  ... \\
33      & $\!\!\!$[0.1] &    $31.11  \pm   0.10$    &     $1.5\pm0.3$         &   $30.53  \pm  0.06$   &      $1.3\pm0.3$            &  0.8  &   $\negsp <0.01 $ &   $\negsp  >0.65 $    &  ...  \\
34      &     0.1   &    $30.81  \pm   0.18$    &     $1.6\pm0.5$         &   $31.64  \pm  0.03$   &      $1.9\pm0.2$            &  0.1  &   $\negsp <0.01 $ &   $\negsp >0.45 $    &  \cite{2003Sci...299.1372S}    \\
35      &     0.25  &    $31.04  \pm   0.10$    &     $3.3\pm0.5$         &   $30.79  \pm  0.13$   &      $1.5\pm0.7$            &  0.2 &
              ... & ...   &  \cite{2003ApJ...585L..41R, 2007ApJ...654..487N} \\
36      &     0.6   &    $31.19  \pm   0.03$    &     $1.5\pm0.1$         &   $30.56  \pm  0.11$   &      $1.1^{+0.4}_{-0.2}$    &  0.1 &   $\negsp <0.005$ &   $\negsp  >
                                       0.5$    &  \cite{2006ApJ...647.1300M}    \\
37      &     0.03  &    $29.11  \pm   0.07$    &     $1.0\pm0.2$         &   $30.32  \pm  0.04$   &      $1.56\pm0.24$          &  0.02 &   $\negsp <0.001$ &   $\negsp >
                                      9$    &  \cite{2006ApJ...643.1146P}    \\
38      &     0.1   &             ...           &         ...             &   $29.82\pm0.10$          &      $2.6\pm0.2$            &  ...  &     ...     &     ...         &   \cite{2006AnA...448L..13H}    \\
39      &     0.17  &    $29.63  \pm   0.01$    &     $1.7\pm0.6$         &   $30.15  \pm  0.06$   &      $1.7^{+0.5}_{-0.7}$    &  0.05 &   $\negsp <0.002$ &   $\negsp  >
                                       0.3$    &  \cite{2006ApJ...645.1421B, 2007arXiv0711.4171M} \\
40      &     0.2   &             ...           &         ...             &   $30.18    \pm 0.11$    &      $2.0\pm0.3$            &  ...  &     ...     &   $   ...  $    &  \cite{2007AandA...467.1209H}    \\
\hline
\end{tabular}
\caption{
X-ray properties of the PWNe and their parent pulsars listed in Table \ref{tab:prop1}}
\label{tab:prop2}
\end{table}

\begin{table}
\setlength{\tabcolsep}{0.05in} \tabletypesize{\footnotesize}
  {\footnotesize
\begin{tabular}{lcccccccccc}
\hline
\tablehead{1}{c}{b}{   \# \\  }  &
  \tablehead{1}{c}{b}{  ${\rm SNR/PWN}$\tablenote{Superscript $^{\rm T}$ marks PWNe in which the point source (likely a pulsar) shows a thermal component.} \\  }  &
  \tablehead{1}{c}{b}{  $d$\tablenote{Our best guess for
the distance,
used to scale
the distance-dependent
parameters in this table.}  \\ ${\rm kpc}$ }  &
 \tablehead{1}{c}{b}{   $n_{\rm H,22}$\tablenote{Hydrogen column density (in units of $10^{22}$ cm$^{-2}$) obtained from spectral fits to the PWN spectra.} \\  }  &
  \tablehead{1}{c}{b}{   $\log L_{\rm pwn}$\tablenote{Logarithm of unabsorbed PWN luminosity in the 0.5--8 keV band, in units of  ergs
s$^{-1}$.  For the bright, significantly extended PWNe (\#\# 44, 45, 46, 47, 48, 51, 52, 53),  we quote
the luminosity of the PWN core.
For PWNe with very extended tails
(\#\# 47, 51, 52), we quote only the luminosity of the bright ``bullet'' component;
the tail luminosities
are listed in Table \ref{tab:tails}.} \\  }  &
  \tablehead{1}{c}{b}{   $\Gamma_{\rm pwn}$  \\  }  &
  \tablehead{1}{c}{b}{   $\log L_{\rm psr}^{\rm nonth}$~\tablenote{Logarithm  of nonthermal luminosity of the PWN
point source (likely pulsars) in the 0.5--8 keV band, in units of
ergs s$^{-1}$.
} \\  }  &
   \tablehead{1}{c}{b}{   $\Gamma_{\rm psr}$  \\  }  &
     \tablehead{1}{c}{b}{   $l_{X}$\tablenote{
Characteristic size of the PWN core in which the PWN X-ray properties were measured.}  \\  ${\rm pc}$ }  &
\tablehead{1}{c}{b}{   ${\rm Rad./H}_{\alpha}{\rm TeV}$\tablenote{Is the PWN detected in radio/H$_{\alpha}$/TeV?
P = `possibly'. }  \\   }  &
\tablehead{1}{c}{b}{   ${\rm Refs.}$  \\  }
 \\
\hline
41 & G0.9+0.1/G0.87+0.08                     & 10  &  $16\pm2$            &    $35.41\pm0.07$ & $2.3\pm0.4$         &  $\lesssim 32.80$  & ...                                &  2.2  &  Y/N/Y    &  \cite{2001ApJ...556L.107G} \\
42 & G16.7+0.1/G16.73+0.08                  &   10  &  $4.7\pm1.0$         &    $34.37\pm0.07$ & $1.2\pm0.3$         &  $\lesssim 33.95$  & ...                              &  2.2  &  P/N/N    &  \cite{2003ApJ...592..941H}  \\
43 & .../G25.24--0.19                &   10  &  $4\pm1$             &    $33.63\pm0.14$ & $0.8\pm0.3$         &  $34.85\pm0.06$    & $0.6\pm0.1$                      &  2.2  &  P/N/Y    &  ... \\
44 & 3C\,396/G39.22--0.32              &  ~~8   &  $5.3\pm0.9$         &    $34.09\pm0.10$ & $1.5\pm0.2$         &  $\lesssim 33.0$   & ...                          &  1.6  &  P/N/N    &  \cite{2003ApJ...592L..45O}   \\
45 & CTB\,87/G74.94+1.11              &  ~~6   &  $1.4\pm0.3$         &    $33.53\pm0.11$ & $1.6\pm0.2$         &  $\lesssim 32.0$   & ...                           &  3.5  &  P/N/N    &  ... \\
46 & CTA\,1/G119.65+10.46$^{\rm T}$      &   ~~~~1.4 &  $0.28\pm0.6$        &    $31.38\pm0.20$ & $1.1\pm0.6$         &  $30.92\pm0.32$    & $1.6\pm0.6$          &  0.14 &  N/N/N    &  \cite{2004ApJ...612..398H} \\
47 & IC\,443/G189.23+2.90$^{\rm T}$      &   ~~~~1.5 &  $0.72\pm0.6$        &    $32.62\pm0.03$ & $1.7\pm0.1$         &  $30.72\pm0.20$    & $2.6^{+0.5}_{-1.0}$  &  0.3  &  Y/P/N   &  \cite{2006ApJ...648.1037G} \\
48 & MSH\,11--62/G291.02--0.11      &  ~~~~1.0 &  $1.0\pm0.1$         &    $33.52\pm0.05$ & $1.6\pm0.1$         &  $32.58\pm0.11$    & $1.36\pm0.09$                 &  1.1   &  P/N/N    &  \cite{2004IAUS..218..203H} \\
49 & G293.8+0.6/G293.79+0.58$^{\rm T}$      &   ~~2   &  $0.4\pm0.2$         &    $30.90\pm0.26$ & $0.4\pm0.3$         &  $\lesssim 31.3 $  & ...                            &  0.6    &  P/N/N    &  \cite{2003AAS...203.3907O} \\
50 & .../G313.32+0.13              &   ~~5   &  $2.3^{+0.7}_{-0.4}$ &    $33.55\pm0.02$ & $1.7\pm0.1$         &  $32.25\pm0.20$    & $1.8\pm0.3$                    &  1.5    &  P/N/N    &  \cite{2005ApJ...627..904N}  \\
51 & MSH\,15-56/G326.12--1.81      &   ~~4   &  $0.6\pm0.2$         &    $32.51\pm0.09$ & $1.7\pm0.2$         &  $31.85\pm0.11$    & $1.5\pm0.2$                    &  1.0    &  Y/N/N    &  \cite{2002APS..APRN17037P}  \\
52 & G327.1--1.1/G327.15--1.04                  &   ~~7   &  $2.2\pm0.2$         &    $34.26\pm0.07$ & $2.0\pm0.1$         &  $33.81\pm0.24$    & $2.3\pm0.4$            &  1.7  &  P/N/N    &  \cite{2004AAS...205.8405S} \\
53 & G12.8--0.0/G12.82--0.02                  &   ~~~~4.5 &  $10\pm1$            &    $32.90\pm0.25$ & $0.4^{+0.4}_{-0.7}$ &  $33.53\pm0.07$    & $1.3\pm0.3$                  &  2.0  &  N/N/P    &  \cite{2007ApJ...665.1297H}  \\
54 & DA\,495/G65.73+1.18$^{\rm T}$    &   ~~~~1.5 &  $0.31\pm0.15$       &    $31.84\pm0.09$ & $1.8\pm0.1$         &  $\lesssim 31.30$  & ...                        &  0.2  &  P/N/N    &  ... \\
\hline
\end{tabular}
}
\caption{Properties of X-ray PWNe without a known pulsar.}
\label{tab:prop3}
\end{table}

Finally, there is a large group of PWNe (see Fig.\ 4)
which we  cannot credibly classify into any
of the above two categories,
 not only because they are  too  small and faint but
 also because some of them exhibit
very bizarre morphologies (e.g., \#\# 12, 14, 40, 44).
One of the most weird objects is \#40 near the old pulsar B2224+65,
whose X-ray image \cite{2003IAUS..214..135W, 2007AandA...467.1209H} shows
a $2'$-long  linear structure almost perpendicular to the pulsar's velocity
and to the orientation of the H$_\alpha$ ``Guitar nebula''
\cite{2002ApJ...575..407C}.
Our examination of
the currently available data suggests, however,
that the feature does not connect to the pulsar but apparently
originates from a nearby point-like source of an unknown nature.

To conclude, although crude systematization of PWN morphologies is
possible,
they do not perfectly fit into just two types,
indicating that the PWN appearance depend not only on the pulsar's
Mach number but also on other parameters.

  \begin{figure}[h]
\includegraphics[width=.48\textwidth,angle=0]{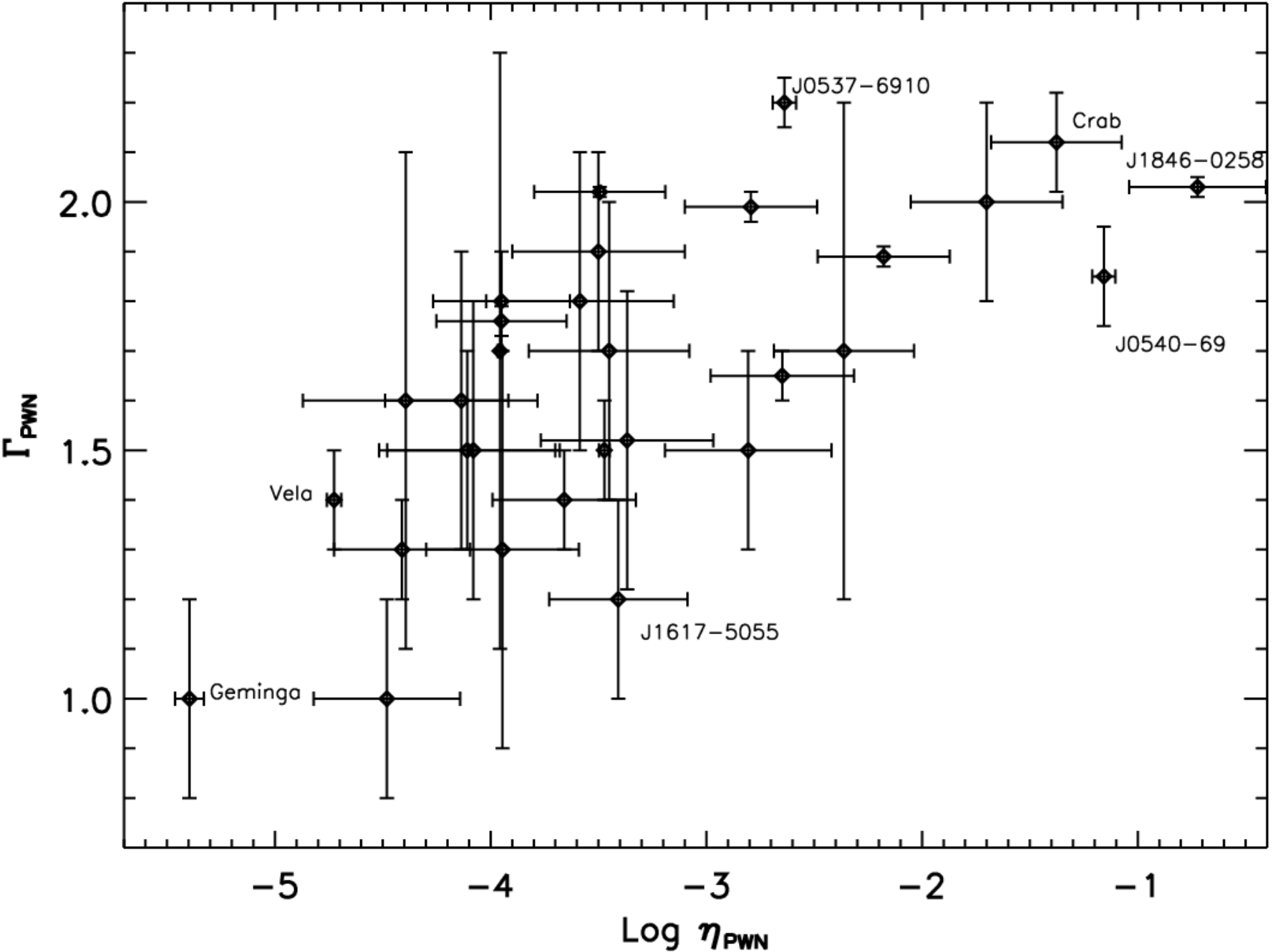}
\caption{Slope of the (spatially averaged) PWN spectrum vs.\ X-ray efficiency.}
\end{figure}

\section{Luminosities and spectra}

 We have
compiled the properties of PWNe
observed with {\sl Chandra} in Tables 1--3.
Even if a PWN had been described previously,
 we reanalyzed the data
to ensure the uniformity
of the analysis;
therefore, our results may differ from
those
 published  by other authors.
Because of the limited space, we discuss here only the luminosities and, very
briefly, the spectra of PWNe, leaving more detailed discussions to a future
paper.

\begin{table}
 \setlength{\tabcolsep}{0.05in}
  {\footnotesize
\begin{tabular}{lrcccccccc}
\hline
   \tablehead{1}{c}{b}{\footnotesize  ${\rm PSR}$ \\  }
  & \tablehead{1}{c}{b}{\footnotesize  $P$  \\  ${\rm ms}$ }
  & \tablehead{1}{c}{b}{\footnotesize  $\log\tau$ \\  }
  & \tablehead{1}{c}{b}{\footnotesize  $\log\dot{E}$
\\  }
  & \tablehead{1}{c}{b}{\footnotesize $d$\tablenote{Our best guess for
the distance to the pulsar, based on the dispersion measure distances \cite{1993ApJ...411..674T, arXiv:astro-ph/0207156C}.
}  \\  ${\rm kpc}$  }
  & \tablehead{1}{c}{b}{ \footnotesize $n_{\rm H,22}$\tablenote{Hydrogen column density  estimated  from the pulsar's dispersion measure (assuming 10\% ISM ionization).} \\  }
  & \tablehead{1}{c}{b}{\footnotesize  ${\rm Exp.}$\tablenote{{\sl Chandra} ACIS exposure time.}
\\  ${\rm ks}$ }
  & \tablehead{1}{c}{b}{\footnotesize  $\log L_X$\tablenote{3$\sigma$ upper limit on
PWN+PSR luminosity in the 0.5--8 keV band,
calculated for an  $R=5''$ circular aperture centered on the radio pulsar position.}
   \\  }
   \\
\hline
 J1105--6107    &  63   &  4.80  &  36.39  & 6    & 0.84 &  11.6  & 31.81 \\
 J1837--0604    &  96   &  4.53  &  36.30  & 6    & 1.43 &  9.1   & 32.05  \\
 J1913+1011   &  36   &  5.23  &  36.46  & 5    & 0.55 &  19.8  & 31.35  \\
 J1906+0746   &  144  &  5.05  &  35.43  & 5   & 0.67 &  32.0  & 31.19  \\
\hline
\end{tabular}
}
\caption{Young pulsars undetected in X-rays.}
\label{tab:nondet}
\end{table}
\normalsize

Figure 5 shows the correlations of
the nonthermal X-ray luminosities of PWNe and pulsars, $\lpwn$ and $\lpsr$,
with the pulsar's
spin-down power $\edot$.
Since the pulsar/PWN X-ray emission
is powered by the
pulsar's spin-down, $\lpwn$ and $\lpsr$ are generally higher for pulsars with larger
$\edot$
\cite{1988ApJ...332..199S, 1997A&A...326..682B, 2002A&A...387..993P, 2004ApJ...617..480C,
2007arXiv0707.4279L}.
However,
both the $\lpwn$-$\edot$ and $\lpsr$-$\edot$ correlations
show huge dispersions.
This is particularly well seen if we compare the X-ray efficiencies,
$\eta_{\rm pwn}=L_{\rm pwn}/\edot$ and $\etapsr=\lpsr/\edot$,
which range from $\sim 10^{-5}$ to
$\sim 10^{-1}$.
Such a large scatter cannot be explained by distance uncertainties
(even though they are poorly constrained for some objects\footnote{
In addition to the statistical uncertainties given in
Tables 2 and 3,
the error bars in Figs.\ 5 and 6
 include
the systematic 40\%  uncertainty ascribed to the distance
when no parallax was measured. }.)
We could explain the scatter of $\lpsr$ by anisotropy of pulsar emission,
which was neglected in the luminosity estimates, but this argument
does not apply to PWN emission which is expected
to be more isotropic. Therefore, we have to conclude
that $\lpsr$ and $\lpwn$ should depend, in addition to $\edot$, on other
parameters, such as the pulsar's magnetic field and the angle between
the magnetic and spin axes.

The highest efficiencies, $\etapwn\approx 0.2d_{19}^2$ and $\etapsr\approx 0.02d_{19}^2$,
where $d_{19}=d/19\,{\rm kpc}$,
are observed for the
PSR/PWN \#12 in the Kes\,75 SNR, which has the smallest spindown age,
$\tau=726$ yr, and the highest magnetic field, $B=4.9\times 10^{13}$ G.
Even if the distance was
somewhat overestimated
\cite{2007arXiv0711.4107L},
the efficiencies are still high enough to speculate that they might be
partly due to some magnetar-type activity
(e.g., the energetic PWN might be generated in a series of magnetar-like
bursts).

Very low efficiencies, $\lesssim 10^{-4}$,
 are observed for the Vela and some of young (Vela-like)
and middle-aged pulsars and their PWNe. Moreover,
 in several cases
neither the pulsar nor its PWN
was detected (see Table 4), including the most X-ray underluminous PSR
J1913+1011, for which $\eta_{\rm pwn} + \eta_{\rm psr} < 8\times 10^{-6}d_{5}^2$.
Since the luminosities of these PSR/PWN pairs can be much lower than
the upper limits,
and possibly there are many other underluminous pulsars/PWNe
that have not been
observed or reported, one cannot derive a reliable correlation law from the
current (biased) sample. We can only crudely estimate upper bounds
in the $L$-$\edot$ correlations,
 $\lpwn \lesssim 10^{33.4}\edot_{36}^{1.6}$ and $\lpsr\lesssim 10^{33.8}\edot_{36}^{1.3}$ ergs s$^{-1}$,
shown by dashed lines in Figure 5.

Remarkably, the correlation between $\lpwn$ and $\lpsr$
(Fig.\ 5, bottom) is much stronger
than between these luminosities and $\edot$. Although this might
be partly due to the fact that $\lpwn/\lpsr$ does not depend on
errors in the distance estimates,
it also indicates that the two luminosities are determined
by the same ``hidden parameters''.
Moreover, the
pulsar and PWN X-ray efficiencies are quite close to each other:
$0.1 \lesssim \etapwn/\etapsr \lesssim 10$,
with an average value $\langle\etapwn/\etapsr\rangle
\sim 4$, for this sample (see also \cite{2007ApJ...660.1413K}).
This result is surprising because
it is hard to expect so similar efficiencies from the
pulsar magnetosphere and the PWN, where the properties of the emitting particles
and the magnetic fields are so different.

For all the objects studied,
the PWN and pulsar nonthermal spectra can be
satisfactorily described by the power-law (PL)
model, with photon indices in the range $1\lesssim \Gamma\lesssim 2$, which corresponds
to the slopes $1\lesssim p\lesssim 3$ for the energy spectrum of emitting particles
($p=2\Gamma -1$).
Although some correlations between the pulsar and PWN spectral slopes,
$\Gamma_{\rm psr}$ and $\Gamma_{\rm pwn}$, have been
previously reported, as well as between the slopes and luminosities
 \cite{2003ApJ...591..361G}, we
 did not find
such correlations at a statistically significant level
in our larger sample.
However, we see a hint of correlation
  between  $\eta_{\rm pwn}$  and $\Gamma_{\rm pwn}$,
such that more X-ray efficient PWNe
may have softer spectra (see
  Fig.\ 6).
Although this can  merely reflect the fact
   that in more luminous but remote  PWNe the
spectral extraction includes regions where the wind particles has cooled radiatively
   and hence
show softer spectra, the comparison of the spatially well-resolved spectra of the Crab
and Vela PWNe supports this correlation.

To conclude,
the larger PWN sample shows a large scatter of PSR/PWN efficiencies,
strong correlation between the pulsar and PWN luminosities, a lack
of strong correlation between spectral slopes, and possible
$\Gamma_{\rm pwn}$-$\etapwn$ correlation.
We should not forget, however, about the large
statistical and systematic
uncertainties of
spectral measurements,
especially those caused by the spatial averaging.
Therefore, {\em deep
observations of a few bright,
 well-resolved PWNe, such as the Crab and Vela,
 currently provide the most efficient way to study
the pulsar winds and their connection to the pulsar properties.}

\section{The Vela PWN}

\begin{figure}[h]
 \centering
\includegraphics[width=0.42\textwidth,angle=0]{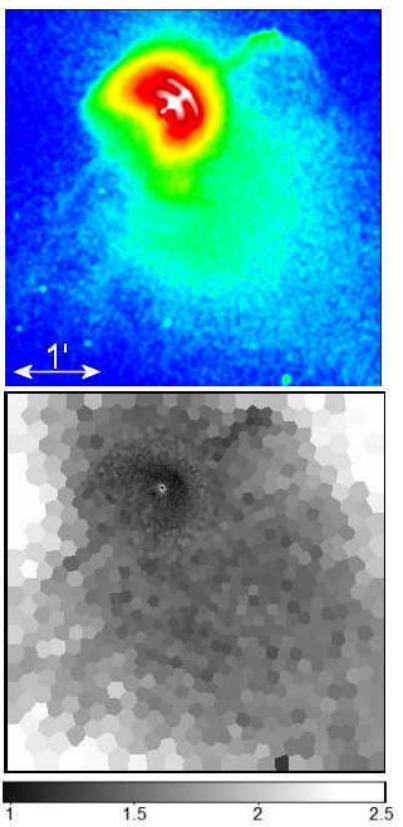}
\caption{Deep image and photon index map of the Vela PWN.
To produce the adaptively binned spectral map,
we used the WVT binnig procedure \cite{2006MNRAS.368..497D}.}
\end{figure}

One of the most interesting among the whole sample
is the Vela PWN generated by the young
($\tau = 11$ kyr), nearby ($d=290$ pc) pulsar B0833--45.
\chan\ observations
have revealed a highly-structured,
bright nebula
\cite{2000AAS...196.3704P, 2001ApJ...556..380H, 2001ApJ...554L.189P, 2003ApJ...591.1157P}
(see \#13 in Figs.\ 2 and 3, and Fig.\ 7).
Its most prominent features
are the
two arcs,
and the northwest (NW) and southeast (SE) ``jets'' along the direction of the pulsar's
proper motion.
The inner arc is certainly a brightened part of an ellipse (a tilted ring),
with the pulsar being offset from the ellipse center, while the topology
of the outer arc
and the  jets
is not so clear.
Combined
images from several observations
reveal fainter structures, such as
  a curved
  extension of the NW jet with brighter ``blobs'',
a fainter extension of the SE jet (Fig.\ 7),
  a puzzling ``bar''
at the apparent origin of the SE jet (see \#13 in Fig.\ 1),
and asymmetric emission extended toward SW, almost perpendicular to the pulsar's
proper motion.
   Although fainter than the arcs and the inner jets, the outer NW jet
  was bright enough
to detect variability of its shape and brightness and measure the speed of blobs,
0.3--0.6\,$c$, moving away from the pulsar
\cite{2003ApJ...591.1157P}.
The inner PWN elements are also
    remarkably variable\footnote{
See a movie at \url{http://www.astro.psu.edu/users/green/pwne/pwne.html#vela}},
 with the outer arc changing its
brightness and curvature and moving back and forth,
and ``knots''
 appearing, disappearing and moving in the dimmer part of the inner ellipse,
SE of the pulsar.
The interpretation of the PWN structure is still debated.
The inner arc (ring) certainly marks a TS, either in an equatorial
outflow \cite{2001ApJ...556..380H}
or in a particle beam precessing around the spin axis \cite{2007ApJ...656.1038D},
and the outer jets are outflows along the spin axis.
However, the nature of the outer
  arc (a ringlike TS offset from the equatorial plane or a convex bowshock-like surface?),
the inner jets (polar outflows or Doppler-boosted images of precessing jets?)
as well as the connection between the inner and outer jets
remains unclear.
The bar at the base of the SE inner jet (a TS in a polar outflow?)
is another piece of the puzzle that remains to be solved.

   Thanks to its proximity,
the compact Vela PWN is both bright and large enough for detailed
   spatial spectroscopy. The adaptively binned  spectral map reveals
strong correlation with the PWN
structure (Fig.\ 7).
Thanks to their hard spectra, both outer
   jets are easily identifiable in the spectral map.  The inner jets and the arcs
also exhibit very hard spectra, $\Gamma$=0.9--1.2,  while the surrounding diffuse
emission is much softer,  $\Gamma\simeq 1.5$. Surprisingly, emission far
SW from the pulsar is relatively hard ($\Gamma=1.3$--1.4);
the X-ray-emitting particles might be supplied there through
the bent outer NW jet ($\Gamma=1.2$--1.3).
Although the physical mechanism responsible for particle
acceleration in pulsar winds is unknown,
    the hard spectrum observed in the Vela PWN,
and the significant differences between the
    spectral slopes in other PWNe\footnote{In particular, the X-ray spectrum of the Crab,
the only other
PWN where the detailed spectral structure has been investigated
\cite{2004ApJ...609..186M}, is much softer;
the photon index measured just downstream of the TS is $\Gamma\approx1.8$, close
that of the Crab pulsar.} (Fig.\ 6), suggest
that
it is not a Fermi-type ultrarelativistic  shock acceleration
        \cite{2001MNRAS.328..393A}
because the latter predicts a universal
slope of the particle spectrum,
$p\approx2.2-2.3$ ($\Gamma\approx 1.60$--1.65).
A possible alternative is acceleration through
heating of ion-electron-positron plasma  by the relativistic
ion-cyclotron instability \cite{2006ApJ...653..325A}, which may
produce spectra with various slopes, depending on the proton-to-pair ratio
in the wind\footnote{This mechanism, however, assumes
a nucleonic component in the pulsar wind, which is yet to be
confirmed observationally.}.

   \begin{figure}[h]
 \centering
\includegraphics[width=0.47\textwidth,angle=0]{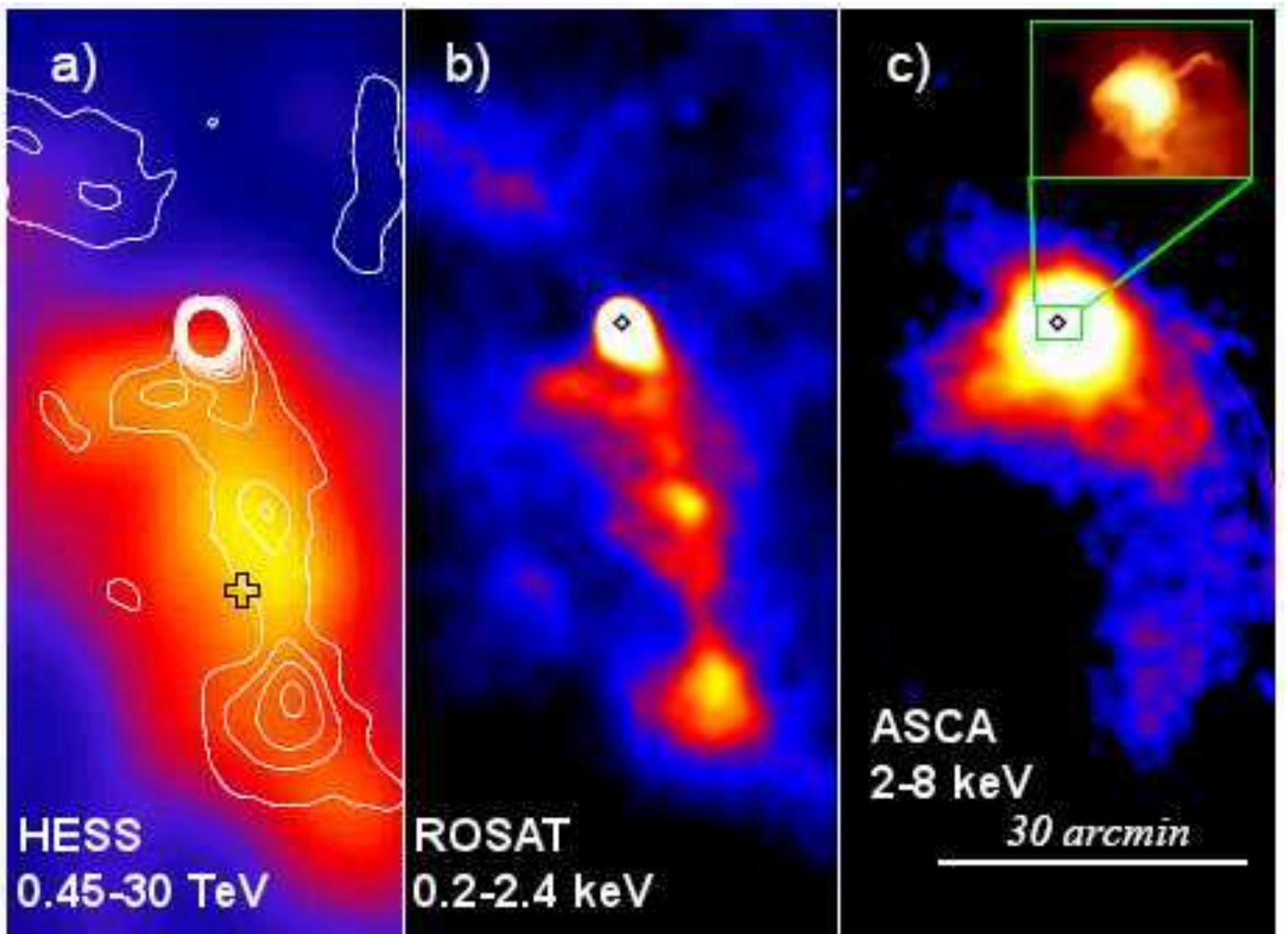}
\caption{Vela X in TeV (left; soft X-ray contours overlayed), soft
X-rays (middle) and harder X-rays (right). Adapted from \cite{2008...Mori}.
}
\end{figure}

The large-scale structure of the Vela PWN has been   studied with
{\sl ROSAT}, {\sl ASCA} and {\sl Suzaku} (\cite{1995Natur.375...40M,
1997ApJ...480L..13M, 2008...Mori}),
revealing an X-ray counterpart to the brightest
filament
in the Vela X radio plerion, presumably created by the Vela pulsar.
The same feature has been recently resolved in the TeV range with HESS
\cite{2006A&A...448L..43A}. The remarkable
  correspondence between the TeV and  X-ray morphologies
(Fig.\ 8) suggests that emission is produced
  by the same particles. However, more detailed investigations show
that the diffuse X-ray emission consists of
a soft thermal component, presumably emitted from the SNR plasma,
 and a hard nothermal component, which is
likely a
large-scale SW extension of the compact PWN
(panels b and c in Fig.\ 8).
Surprisingly, the TeV
  emission is much better correlated with the soft X-ray component
than with the hard one, suggesting that collisions of the pulsar
wind with the SNR matter might play some role in generating the TeV
emission (e.g., by producing $\pi^0$ mesons which decay into TeV photons).
Future multiwavelength observations (e.g., with {\sl GLAST}) may test
this hypothesis.

\section{TeV plerions}

  In addition to promoting Vela X  to the rank of TeV plerions,
HESS observations of the Galactic plane
have revealed
a new population of extended
 TeV sources \cite{2005Sci...307.1938A}.
It has been noticed
(e.g., \cite{2006IAUJD...2E..53D, 2007ApJ...670..643K})
that some
   of
the extended TeV sources
neighbor
 young Vela-like pulsars,
 offset by $10'$--$20'$ from the center of
the TeV emission.  To date,
 young
  pulsars
  have been
  found  in the vicinity of $\sim$10  extended TeV sources
(e.g., \cite{2007Ap&SS.309..197G, 2007arXiv0709.3614C}).
Despite the
seemingly large offsets, the chance
 coincidence probability  is very low, $\lesssim10^{-6}$,
  so that the associations must be taken seriously \cite{2007ApJ...670..643K}.

 \begin{figure}[h]
 \centering
\includegraphics[width=3.1in,angle=0]{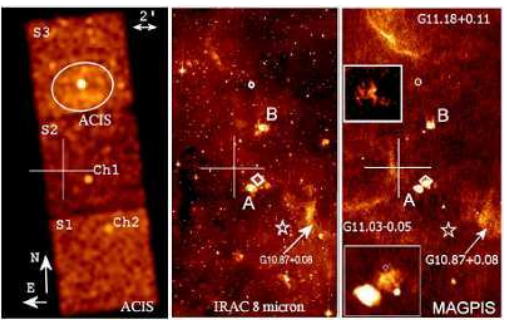}
\caption{Large-scale structure of X-ray (left), IR (middle), and radio
(right) emission around
PSR J1809--1917.
The pulsar and its compact PWN are imaged on the \chan\ ACIS S3 chip.
The center of the TeV source HESS J1809--193 is shown by a cross.
Adapted from \cite{2007ApJ...670..655K}. }
\end{figure}

   Furthermore,  X-ray
  observations  of
   the Vela \cite{2008...Mori},
     PSR~J1826--1334 \cite{2003ApJ...588..441G, 2007arXiv0707.3529P},
 and PSR~J1809--1917 \cite{2007ApJ...670..655K}
 have provided evidence that
 the TeV sources are connected to the pulsars through  faint,
  asymmetric X-ray nebulae.
For example, the elongated morphology of the compact PWN J1809--1917 (\#26 in Fig.\ 3),
with a tail directed to the north from the pulsar,
shows that, similar to the Vela and J1826--1334,
the offset TeV source is not located behind the moving pulsar.
However,  heavily binned X-ray
   images (Fig.\ 9)
     show
     that
the compact PWN is
immersed in an asymmetric large-scale nebula extending
toward the center of the TeV emission.
Such offsets and the asymmetries of the large-scale X-ray PWNe
could be created by the reverse SNR shock that had propagated
through the nonhomogeneous SNR interior and reached one side of
the PWN sooner than the other side \cite{2001ApJ...563..806B}. It has been proposed
 that
 this scenario could also account for the similarly asymmetric, offset TeV
 emission (e.g., \cite{2006IAUJD...2E..53D}).
 However,
the physical origin of the TeV emission still remains unclear.
 It can be produced via the
IC scattering
of seed photons (e.g., the CMB radiation and/or IR background, e.g.,
from nearby star-forming regions and warm dust clouds,
such as sources A and B in Fig.\ 9)
off the relativistic pulsar-wind electrons.
 Alternatively, the TeV photons can be produced
 as a
result of $\pi^{0}\rightarrow\gamma + \gamma$ decay, with $\pi^{0}$
being produced
 when the relativistic protons from the pulsar wind interact with the
 ambient matter \cite{2007Ap&SS.309..189H}.
 If confirmed, the latter mechanism
  may provide the long-sought observational evidence for
the elusive proton component in the
pulsar wind \cite{2007arXiv0708.1050A}.

\footnotesize
\begin{table}
 \setlength{\tabcolsep}{0.06in} \tabletypesize{\footnotesize}
\begin{tabular}{rlclccc}
\hline
   \tablehead{1}{c}{b}{
\# \\  }
  & \tablehead{1}{c}{b}{  $d$\tablenote{
Best-guess distance
used to scale
the distance-dependent
parameters in this table. Distances marked by $^{\rm p}$ are from parallax measurements.}  \\  ${\rm kpc}$ }
  & \tablehead{1}{c}{b}{  $V_{\perp}$\tablenote{Measured (when uncertainties are provided)
or estimated transverse pulsar velocity.} \\ ${\rm km\,\,s}^{-1}$
}
  & \tablehead{1}{c}{b}{  $l_{\rm ext}$\tablenote{
Largest linear extent of the X-ray PWN.
} \\ ${\rm pc}$ }
  & \tablehead{1}{c}{b}{ $\log L_{\rm ext}$\tablenote{Logarithm of luminosity of the
large-scale  PWN component,
in the 0.5--8 keV band.
}  \\   }
& \tablehead{1}{c}{b}{$\log\eta_{\rm ext}$\tablenote{$\eta_{\rm ext}=L_{\rm ext}/\edot$} \\ }
   \\
\hline
1  & $\!\!\!\!$50              &  $\sim600$               &     3.7   &   $36.21\pm0.01$ & $-2.47$ \\
15    &  2.5             &  $280\pm80$              &     1.2   &   $33.02\pm0.11$ & $-3.55$ \\
17   &  2               &  $\lesssim100$             &     3     &   $32.60\pm0.10$ & $-3.93$ \\
22  &  5               &  $\sim 500$                 &     1.1   &   $33.83\pm0.09$ & $-2.57$ \\
29  &  4               &  $300-900$                  &     6.5   &   $33.05\pm0.04$ & $-2.66$ \\
30     &  3               &  $300-400$               &     1.3   &   $32.20\pm0.10$ & $-2.58$  \\
33 &  1.4             &  $200-1000$                  &     2     &   $31.60\pm0.10$ & $-3.76$  \\
35 &  $1.47^{\rm p}$  &  $400^{+114}_{-73}$        &     1     &   $31.30\pm0.15$ & $-3.39$ \\
36   &  $1.04^{\rm p}$  &  $61^{+12}_{-9}$          &     1.5   &   $31.20\pm0.07$ & $-3.46$  \\
37 &  $0.25^{\rm p}$  &  $211\pm2$                &     0.35  &   $29.35\pm0.11$ & $-5.53$ \\
38 &  0.25            &  $\sim56\pm4$               &     0.04  &   $28.98\pm0.15$ & $-4.85$ \\
39   &  $0.36^{\rm p}$  &  $177^{+4}_{-5}$          &     1.5   &   $29.50\pm0.25$ & $-4.09$ \\
40   &  1               &  $862\pm14$                &     0.6   &   $30.18\pm0.10$ & $-2.89$  \\
47          &  1.4             &  $\sim250$          &     0.65  &   $32.82\pm0.03$ & ... \\
51 &  4               &  $100-400$                  &     3.5   &   $32.8\pm0.2$ & ... \\
52 &  7               &  $\sim500$                   &     5.6   &   $33.09\pm0.10$ & ... \\
\hline
\end{tabular}
\caption{Properties of  large-scale components in ram-pressure confined  PWNe.}
\label{tab:tails}
\end{table}
\normalsize

\section{Long Pulsar Tails}

\chan\ observations have allowed us to discover several long, up to a few parsecs,
pulsar tails,
obviously associated with the pulsar motion (see Fig.\ 3
and Table 5).
 The longest X-ray tail ($l=6.5$ pc at $d=4$ kpc, limited by the detector FOV)
 was observed behind PSR J1509--5058 (hereafter J1509; see Fig.\ 10).
The large lengths of this and other tails
indicate that they are not just trails of ``dumped'' electrons (or positrons)
behind the moving pulsars because
such an assumption would imply an improbably high pulsar speed:
$v_{\rm psr}=l/\tau_{\rm syn}\sim
10,000
 (l/6\,{\rm pc})
(B/10\,\mu{\rm G})^{3/2}(E/3\,{\rm keV})^{1/2}$ km s$^{-1}$,
where $\tau_{\rm syn}$ is the synchrotron cooling time for electrons
that emit photons of energy $E$.
Therefore, the tails represent
ram-pressure confined streams of relativistic electrons
with a large bulk flow velocity, $v_{\rm flow}\gg v_{\rm psr}$.
This conclusion is also supported by the detection of the exremely long
(17 pc at $d=5$ kpc) radio tail of the Mouse PWN \cite{2005AdSpR..35.1129Y}.
On the other hand, if one assumes a nearly relativistic flow velocity, such
as obtained in
 the numerical
simulations
\cite{2005A&A...434..189B}
for the ram-pressure confined flow
just behind the back surface of the TS bullet,
then the tail
should be much longer than observed.
This suggests that the flow decelerates on a length scale of $\lesssim$1 pc
(time scale $\lesssim$10--100 yrs), perhaps due to
shear instabilities at the CD surface and entrainment of the ambient
matter.
Measuring the spectral
changes along the tail and confronting them with the models of cooling
MHD flows
can constrain the evolution of the
flow speed along the tail and elucidate
the deceleration
mechanism(s).

  \begin{figure}[h]
 \centering
\includegraphics[width=.42\textwidth,angle=0]{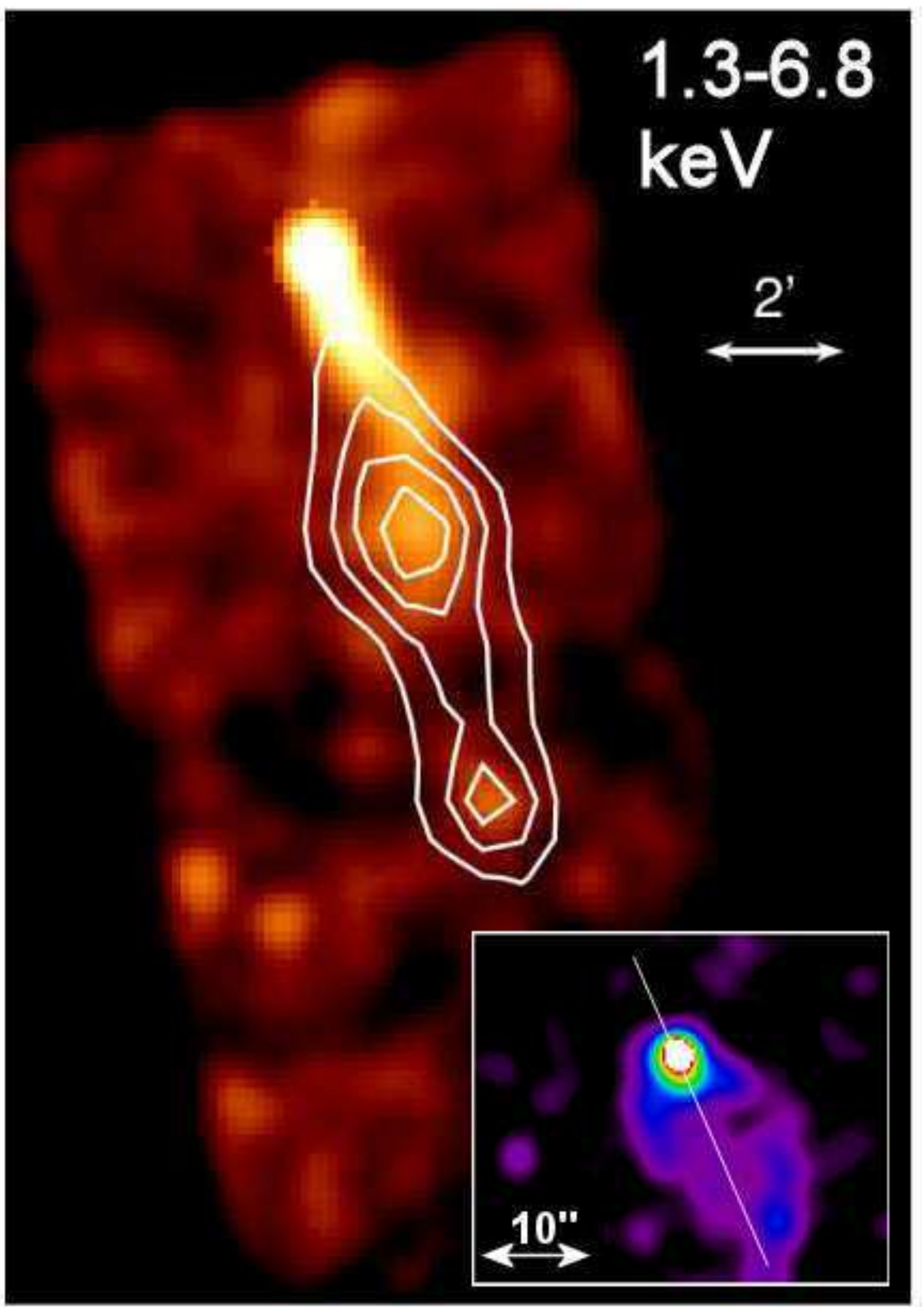}
\caption{X-ray tail of J1509--5058 with 843-MHz contours overlayed (Kargaltsev et al., in prep.).}
\end{figure}

If the tail flow
indeed starts as mildly relativistic, then the tail's properties
should be
very similar to those of
a pulsar jet. Moreover, if a pulsar moves along its spin axis and the wind outflow
has polar components (this scenario has not been considered in
the numerical simulations), then the ``tail''
{\em is} actually a ``rear jet'' (the front jet may be destroyed by the ram pressure
or not seen because of the Doppler effect).
In this case, we can expect kink and sausage instabilities in the pinched
flow, similar to those responsible for bends and blobs
 in the Vela outer jet \cite{2003ApJ...591.1157P}.
The nonuniformities observed
in the tails of J1509 and PSR B1929+10 \cite{2007arXiv0711.4171M} might be a manifestation of
such instabilities, which could be used to measure the flow velocity in a series
of deep observations.

Despite its lower surface brightness, the luminosity of the extended tail of J1509,
$L_{\rm tail}\approx 1.1\times 10^{33}$ erg s$^{-1}$
($\eta_{\rm tail}\approx2\times10^{-3}$),
 exceeds
the luminosity of the compact ($\sim20''\times 8''$) PWN ``head''
by a factor
of 10.
This example demonstrates that
   pulsar tails may appear more luminous than
compact PWNe
and explains the ubiquity of tails
found behind pulsars with
lower $\dot{E}$ (e.g., \# 29 through \#40 in Table 1).
The reason is that in a ram-pressure confined PWN
the entire wind flow is channeled into a
narrow linear structure, which results in
a larger column density of wind particles
(hence higher surface brightness) at a given distance from the pulsar
and makes it easier to detect the PWN
emission
much further away from the pulsar than
in the case of a more isotropic PWN.
Thus, spatially resolved X-ray spectroscopy and multiwavelength observations of
pulsar tails
are most useful for studying the evolution of cooling pulsar winds
and the properties of relativistic MHD flows.
 Particularly important are radio observations
as they can map the magnetic
field
within the tail through the polarization measurements.

\section{Conclusions}

The high angular resolutions and sensitivity of \chan\ have allowed us to
detect many X-ray PWNe and study their structure and spectra. Our current
understanding of X-ray PWNe can be briefly summarized as follows.

\begin{itemize}
\item Most of the detected PWNe are
associated with young, powerful pulsars (partly due to
selection effect), but some old PWNe have also been detected (e.g., the
tail of PSR B1929+10).

\item The observed PWN morphologies can be crudely classified into
the torus-jet and bowshock-tail types, corresponding to sub- and supersonic
pulsar motion, respectively. However, the classification is often uncertain,
and many morphological features (e.g., in the Vela and Geminga PWNe) remain
to be understood.

\item PWNe radiate up to a few percent of the pulsar spin-down
power
in X-rays.
The X-ray PWN efficiency generally
 grows with $\edot$, but the $\etapwn$-$\edot$ dependence shows a very large
scatter, including some very dim PWNe, $\etapwn < 10^{-5}$.

\item The X-ray
luminosities of the detected PWNe and their parent pulsars
are
of the same order of magnitude, with PWNe being, on average, somewhat more
luminous.

\item The photon indices of the PWN spectra are in the range
$1\lesssim \Gamma \lesssim 2$. More X-ray efficient PWNe apparently show
softer spectra, but this conclusion should be checked in deeper observations
of well-resolved PWNe to mitigate the effects of synchrotron cooling.

\item Spectral maps of the Crab and Vela PWN show strong correlation between
the spectral hardness and morphological features. The Crab's spectra immediately downstream
of TS are considerably softer than those of the Vela.

\item X-ray observations of
pulsars/PWNe in the vicinity of
extended TeV sources
show faint extensions of compact PWNe toward
the offset centers of the TeV emission, supporting the interpretation of
these TeV sources as ``crushed plerions''.

\item Parsec-scale
X-ray tails found behind many pulsars represent
ram-pressure collimated, high-velocity flows of relativistic particles,
resembling pulsar jets in their properties.
\end{itemize}

Notwithstanding the considerable progress in
observations and modeling of PWNe,
there still remains a nubmer of important problems to solve.
Here we mention a few of them.

\begin{itemize}
\item What is the origin
of the diverse,
often very complex, PWN morphologies? For instance,
what is the nature of the outer arc, inner jets, and bar in the Vela PWN,
the loops in the 3C\,58 PWN, the shell and axial tail in the
Geminga PWN?

\item Where and how the X-ray emitting particles are accelerated?
Does the acceleration occur in the preshock wind or at the TS?
Are the winds composed of electrons/positrons or they have some
nucleonic component?

\item What are the physical parameters that determine the X-ray PWN
efficiency, in addition to $\edot$?
Why don't we see any PWN around some pulsars? Is it because of
some special properties of the pulsar wind or the ambient medium?

\item Why are the particle spectra so hard in some PWNe (e.g., Vela) and
much softer in others (e.g., Crab)?
Is such a difference
caused by different properties of the parent pulsars'
winds (e.g.,
magnetization),
or different
efficiencies of particle acceleration between the magnetosphere and the TS?
What is the reason for the apparent efficiency-hardness correlation?

\item Can all the Galactic extended TeV sources
with luminous pulsars nearby be interpreted as relic TeV plerions
or there are indeed ``dark accelerators''? Are these sources indeed ``crushed
PWNe'' formed by the passage of the inverse SNR shock? Can the huge sizes
of the TeV plerions and the large offsets from the pulsars
 be reconciled with the ``crushed PWN'' scenario?
Is the TeV radiation due to the IC scattering or the pion decay? If the former,
are the seed photons supplied by the CMB or IR radiation from nearby star-forming
regions or dust clouds?

\item What are the flow velocities and magnetic fields in the long pulsar tails?
Which role does the magnetic field play in the tail collimation and what is the magnetic field  topology ?
How do these collimated flows decelerate and cool?
\end{itemize}

To answer these questions,
new observations are badly needed, both in X-rays and other wavelengths, as well as
theoretical work and numerical modeling. Particularly important would be to
take full advantage of the outstanding \chan\ capabilities as long as it is alive
because no X-ray observatory with such high angular resolution is expected
in the foreseeable future.

\begin{theacknowledgments}
 We thank Koji Mori for providing the {\sl ASCA} images of Vela
 X. 
This work was partially supported by Chandra award AR5-606X.
\end{theacknowledgments}




\bibliographystyle{aipprocl} 


\bibliography{bibliograhpy4}

\hyphenation{Post-Script Sprin-ger}
\begin{thebibliography}{10}
\providecommand{\enquote}[1]{``#1''}
\expandafter\ifx\csname url\endcsname\relax
  \def\url#1{\texttt{#1}}\fi
\expandafter\ifx\csname urlprefix\endcsname\relax\def\urlprefix{URL }\fi

\bibitem{1984ApJ...283..710K}
C.~F. {Kennel}, and F.~V. {Coroniti}, \emph{\apj} \textbf{283}, 710--730
  (1984).

\bibitem{2003A&A...404..939V}
E.~{van der Swaluw}, \emph{\aap} \textbf{404}, 939--947 (2003).

\bibitem{2005A&A...434..189B}
N.~{Bucciantini}, E.~{Amato}, and L.~{Del Zanna}, \emph{\aap} \textbf{434},
  189--199 (2005).

\bibitem{2002ApJ...575..407C}
S.~{Chatterjee}, and J.~M. {Cordes}, \emph{\apj} \textbf{575}, 407--418 (2002).

\bibitem{2005AJ....129.1993M}
R.~N. {Manchester}, G.~B. {Hobbs}, A.~{Teoh}, et~al., \emph{\aj} \textbf{129},
  1993--2006 (2005).

\bibitem{2004MNRAS.349..779K}
S.~S. {.Komissarov}, and Y.~E. {Lyubarsky}, \emph{\mnras} \textbf{349},
  779--792 (2004).

\bibitem{2006A&A...453..621D}
L.~{Del Zanna}, D.~{Volpi}, E.~{Amato}, et~al., \emph{\aap} \textbf{453},
  621--633 (2006).

\bibitem{2003A&A...397..913V}
E.~{van der Swaluw}, A.~{Achterberg}, Y.~A. {Gallant}, et~al., \emph{\aap}
  \textbf{397}, 913--920 (2003).

\bibitem{2006csxs.book..279K}
V.~M. {Kaspi}, M.~S.~E. {Roberts}, and A.~K. {Harding}, \enquote{{Isolated
  Neutron Stars},} in \emph{Compact Stellar X-ray Sources}, edited by W.~Lewin,
  and M.~van~der Klis, Cambridge Univ.\ Press, Cambridge, UK, 2006, pp.
  279--339.

\bibitem{2006ARA&A..44...17G}
B.~M. {Gaensler}, and P.~O. {Slane}, \emph{\araa} \textbf{44}, 17--47 (2006).

\bibitem{2000ApJ...536L..81W}
M.~C. {Weisskopf}, J.~J. {Hester}, A.~F. {Tennant}, et~al., \emph{\apjl}
  \textbf{536}, L81--L84 (2000).

\bibitem{1998Natur.393..139S}
H.~C. {Spruit}, and E.~S. {Phinney}, \emph{\nat} \textbf{393}, 139--141 (1998).

\bibitem{2002ApJ...577L..49H}
J.~J. {Hester}, K.~{Mori}, D.~{Burrows}, et~al., \emph{\apjl} \textbf{577},
  L49--L52 (2002).

\bibitem{2004IAUS..218..181M}
K.~{Mori}, D.~N. {Burrows}, G.~G. {Pavlov}, et~al., \enquote{{Year-scale
  Morphological Variations of the X-ray Crab Nebula},} in \emph{Young Neutron
  Stars and Their Environments}, edited by F.~{Camilo}, and B.~M. {Gaensler},
  2004, vol. 218 of \emph{IAU Symposium}, pp. 181--184.

\bibitem{2004ApJ...616..403S}
P.~{Slane}, D.~J. {Helfand}, E.~{van der Swaluw}, et~al., \emph{\apj}
  \textbf{616}, 403--413 (2004).

\bibitem{2001ApJ...556..380H}
D.~J. {Helfand}, E.~V. {Gotthelf}, and J.~P. {Halpern}, \emph{\apj}
  \textbf{556}, 380--391 (2001).

\bibitem{2003ApJ...591.1157P}
G.~G. {Pavlov}, M.~A. {Teter}, O.~{Kargaltsev}, et~al., \emph{\apj}
  \textbf{591}, 1157--1171 (2003).

\bibitem{2002ApJ...569..878G}
B.~M. {Gaensler}, J.~{Arons}, V.~M. {Kaspi}, et~al., \emph{\apj} \textbf{569},
  878--893 (2002).

\bibitem{2001ApJ...559..275W}
Q.~D. {Wang}, E.~V. {Gotthelf}, Y.-H. {Chu}, et~al., \emph{\apj} \textbf{559},
  275--281 (2001).

\bibitem{2006ApJ...651..237C}
Y.~{Chen}, Q.~D. {Wang}, E.~V. {Gotthelf}, et~al., \emph{\apj} \textbf{651},
  237--249 (2006).

\bibitem{2004ApJ...616..383G}
B.~M. {Gaensler}, E.~{van der Swaluw}, F.~{Camilo}, et~al., \emph{\apj}
  \textbf{616}, 383--402 (2004).

\bibitem{2005AdSpR..35.1129Y}
F.~{Yusef-Zadeh}, and B.~M. {Gaensler}, \emph{Adv.\ Space Res.} \textbf{35},
  1129--1136 (2005).

\bibitem{2004ApJ...612..389H}
J.~W.~T. {Hessels}, M.~S.~E. {Roberts}, S.~M. {Ransom}, et~al., \emph{\apj}
  \textbf{612}, 389--397 (2004).

\bibitem{2006ApJ...647.1300M}
K.~E. {McGowan}, W.~T. {Vestrand}, J.~A. {Kennea}, et~al., \emph{\apj}
  \textbf{647}, 1300--1308 (2006).

\bibitem{2003Sci...301.1345C}
P.~A. {Caraveo}, G.~F. {Bignami}, A.~{De Luca}, et~al., \emph{Science}
  \textbf{301}, 1345--1348 (2003).

\bibitem{2006ApJ...643.1146P}
G.~G. {Pavlov}, D.~{Sanwal}, and V.~E. {Zavlin}, \emph{\apj} \textbf{643},
  1146--1150 (2006).

\bibitem{2002ApJ...580L.137G}
B.~M. {Gaensler}, D.~H. {Jones}, and B.~W. {Stappers}, \emph{\apjl}
  \textbf{580}, L137--L141 (2002).

\bibitem{2005AAS...20718313C}
S.~{Chatterjee}, B.~M. {Gaensler}, M.~{Vigelius}, et~al., \emph{BAAS}
  \textbf{37}, 1470 (2005).

\bibitem{2006AnA...448L..13H}
C.~Y. {Hui}, and W.~{Becker}, \emph{\aap} \textbf{448}, L13--L17 (2006).

\bibitem{2004ApJ...609..186M}
K.~{Mori}, D.~N. {Burrows}, J.~J. {Hester}, et~al., \emph{\apj} \textbf{609},
  186--193 (2004).

\bibitem{2001AnA...365L.212W}
R.~{Willingale}, B.~{Aschenbach}, R.~G. {Griffiths}, et~al., \emph{\aap}
  \textbf{365}, L212--L217 (2001).

\bibitem{2001ApJ...546.1159K}
P.~{Kaaret}, H.~L. {Marshall}, T.~L. {Aldcroft}, et~al., \emph{\apj}
  \textbf{546}, 1159--1167 (2001).

\bibitem{2001ApJ...561..308S}
S.~{Safi-Harb}, I.~M. {Harrus}, R.~{Petre}, et~al., \emph{\apj} \textbf{561},
  308--320 (2001).

\bibitem{2001ApJ...552L.125H}
J.~P. {Halpern}, F.~{Camilo}, E.~V. {Gotthelf}, et~al., \emph{\apjl}
  \textbf{552}, L125--L128 (2001).

\bibitem{2001ApJ...559L.153H}
J.~P. {Hughes}, P.~O. {Slane}, D.~N. {Burrows}, et~al., \emph{\apjl}
  \textbf{559}, L153--L156 (2001).

\bibitem{2002ApJ...568L..49L}
F.~J. {Lu}, Q.~D. {Wang}, B.~{Aschenbach}, et~al., \emph{\apjl} \textbf{568},
  L49--L52 (2002).

\bibitem{2005ApJ...627..904N}
C.-Y. {Ng}, M.~S.~E. {Roberts}, and R.~W. {Romani}, \emph{\apj} \textbf{627},
  904--909 (2005).

\bibitem{2003ApJ...582..783H}
D.~J. {Helfand}, B.~F. {Collins}, and E.~V. {Gotthelf}, \emph{\apj}
  \textbf{582}, 783--792 (2003).

\bibitem{2001ApJ...554L.189P}
G.~G. {Pavlov}, O.~Y. {Kargaltsev}, D.~{Sanwal}, et~al., \emph{\apjl}
  \textbf{554}, L189--L192 (2001).

\bibitem{2003ApJ...588..992R}
M.~S.~E. {Roberts}, C.~R. {Tam}, V.~M. {Kaspi}, et~al., \emph{\apj}
  \textbf{588}, 992--1002 (2003).

\bibitem{2004ApJ...610L..33M}
D.-S. {Moon}, J.-J. {Lee}, S.~S. {Eikenberry}, et~al., \emph{\apjl}
  \textbf{610}, L33--L36 (2004).

\bibitem{2005ApJ...628..931L}
X.~H. {Li}, F.~J. {Lu}, and T.~P. {Li}, \emph{\apj} \textbf{628}, 931--937
  (2005).

\bibitem{2005ApJ...631..480R}
R.~W. {Romani}, C.-Y. {Ng}, R.~{Dodson}, et~al., \emph{\apj} \textbf{631},
  480--487 (2005).

\bibitem{2007ApJ...665L.143Z}
V.~E. {Zavlin}, \emph{\apjl} \textbf{665}, L143--L146 (2007).

\bibitem{2003ApJ...588..441G}
B.~M. {Gaensler}, N.~S. {Schulz}, V.~M. {Kaspi}, et~al., \emph{\apj}
  \textbf{588}, 441--451 (2003).

\bibitem{2007arXiv0707.3529P}
G.~G. {Pavlov}, O.~{Kargaltsev}, and W.~F. {Brisken}, \emph{arXiv:0707.3529}
  (2007).

\bibitem{2001ApJ...562L.163K}
V.~M. {Kaspi}, E.~V. {Gotthelf}, B.~M. {Gaensler}, et~al., \emph{\apjl}
  \textbf{562}, L163--L166 (2001).

\bibitem{2004ApJ...616.1118C}
F.~{Camilo}, B.~M. {Gaensler}, E.~V. {Gotthelf}, et~al., \emph{\apj}
  \textbf{616}, 1118--1123 (2004).

\bibitem{2003ApJ...591L.143G}
M.~{Gonzalez}, and S.~{Safi-Harb}, \emph{\apjl} \textbf{591}, L143--L146
  (2003).

\bibitem{2007ApJ...660.1413K}
O.~{Kargaltsev}, G.~G. {Pavlov}, and G.~P. {Garmire}, \emph{\apj} \textbf{660},
  1413--1423 (2007).

\bibitem{2006ApJ...652..569G}
M.~E. {Gonzalez}, V.~M. {Kaspi}, M.~J. {Pivovaroff}, et~al., \emph{\apj}
  \textbf{652}, 569--575 (2006).

\bibitem{2007ApJ...670..655K}
O.~{Kargaltsev}, and G.~G. {Pavlov}, \emph{\apj} \textbf{670}, 655--667 (2007).

\bibitem{2007arXiv0710.0367H}
J.~A. {Hinton}, S.~{Funk}, S.~{Carrigan}, et~al., \emph{arXiv:0710.0367}
  (2007).

\bibitem{2007AandA...470..965H}
C.~Y. {Hui}, and W.~{Becker}, \emph{\aap} \textbf{470}, 965--968 (2007).

\bibitem{2002ApJ...579..404P}
R.~{Petre}, K.~D. {Kuntz}, and R.~L. {Shelton}, \emph{\apj} \textbf{579},
  404--410 (2002).

\bibitem{2003Sci...299.1372S}
B.~W. {Stappers}, B.~M. {Gaensler}, V.~M. {Kaspi}, et~al., \emph{Science}
  \textbf{299}, 1372--1374 (2003).

\bibitem{2003ApJ...585L..41R}
R.~W. {Romani}, and C.-Y. {Ng}, \emph{\apjl} \textbf{585}, L41--L44 (2003).

\bibitem{2007ApJ...654..487N}
C.-Y. {Ng}, R.~W. {Romani}, W.~F. {Brisken}, et~al., \emph{\apj} \textbf{654},
  487--493 (2007).

\bibitem{2006ApJ...645.1421B}
W.~{Becker}, M.~{Kramer}, A.~{Jessner}, et~al., \emph{\apj} \textbf{645},
  1421--1435 (2006).

\bibitem{2007arXiv0711.4171M}
Z.~{Misanovic}, G.~G. {Pavlov}, and G.~P. {Garmire}, \emph{arXiv:0711.4171}
  (2007).

\bibitem{2007AandA...467.1209H}
C.~Y. {Hui}, and W.~{Becker}, \emph{\aap} \textbf{467}, 1209--1214 (2007).

\bibitem{2001ApJ...556L.107G}
B.~M. {Gaensler}, M.~J. {Pivovaroff}, and G.~P. {Garmire}, \emph{\apjl}
  \textbf{556}, L107--L111 (2001).

\bibitem{2003ApJ...592..941H}
D.~J. {Helfand}, M.~A. {Ag{\"u}eros}, and E.~V. {Gotthelf}, \emph{\apj}
  \textbf{592}, 941--946 (2003).

\bibitem{2003ApJ...592L..45O}
C.~M. {Olbert}, J.~W. {Keohane}, K.~A. {Arnaud}, et~al., \emph{\apjl}
  \textbf{592}, L45--L48 (2003).

\bibitem{2004ApJ...612..398H}
J.~P. {Halpern}, E.~V. {Gotthelf}, F.~{Camilo}, et~al., \emph{\apj}
  \textbf{612}, 398--407 (2004).

\bibitem{2006ApJ...648.1037G}
B.~M. {Gaensler}, S.~{Chatterjee}, P.~O. {Slane}, et~al., \emph{\apj}
  \textbf{648}, 1037--1042 (2006).

\bibitem{2004IAUS..218..203H}
I.~{Harrus}, J.~P. {Bernstein}, P.~O. {Slane}, et~al., \enquote{{Chandra Study
  of the Central Object Associated with the Supernova Remnant MSH 11-62},} in
  \emph{Young Neutron Stars and Their Environments}, edited by F.~{Camilo}, and
  B.~M. {Gaensler}, 2004, vol. 218 of \emph{IAU Symposium}, p. 203.

\bibitem{2003AAS...203.3907O}
C.~M. {Olbert}, J.~W. {Keohane}, and E.~V. {Gotthelf}, \emph{BAAS} \textbf{35},
  1265 (2003).

\bibitem{2002APS..APRN17037P}
P.~P. {Plucinsky}, J.~R. {Dickel}, P.~O. {Slane}, et~al., \emph{APS Meeting
  Abstracts,} p. 17037 (2002).

\bibitem{2004AAS...205.8405S}
P.~{Slane}, B.~M. {Gaensler}, E.~{van der Swaluw}, J.~P. {Hughes}, and J.~A.
  {Jenkins}, \emph{BAAS} \textbf{36}, 1481 (2004).

\bibitem{2007ApJ...665.1297H}
D.~J. {Helfand}, E.~V. {Gotthelf}, J.~P. {Halpern}, et~al., \emph{\apj}
  \textbf{665}, 1297--1303 (2007).

\bibitem{2003IAUS..214..135W}
D.~S. {Wong}, J.~M. {Cordes}, S.~{Chatterjee}, et~al., \enquote{{Chandra
  Observations of the Guitar Nebula},} in \emph{High Energy Processes and
  Phenomena in Astrophysics}, edited by X.~D. {Li}, V.~{Trimble}, and Z.~R.
  {Wang}, 2003, vol. 214 of \emph{IAU Symposium}, p. 135.

\bibitem{1993ApJ...411..674T}
J.~H. {Taylor}, and J.~M. {Cordes}, \emph{\apj} \textbf{411}, 674--684 (1993).

\bibitem{arXiv:astro-ph/0207156C}
J.~M. {Cordes}, and T.~J.~W. {Lazio}, \emph{arXiv:astro-ph/0207156}  (2003).

\bibitem{1988ApJ...332..199S}
F.~D. {Seward}, and Z.-R. {Wang}, \emph{\apj} \textbf{332}, 199--205 (1988).

\bibitem{1997A&A...326..682B}
W.~{Becker}, and J.~{Tr\"{u}mper}, \emph{\aap} \textbf{326}, 682--691 (1997).

\bibitem{2002A&A...387..993P}
A.~{Possenti}, R.~{Cerutti}, M.~{Colpi}, et~al., \emph{\aap} \textbf{387},
  993--1002 (2002).

\bibitem{2004ApJ...617..480C}
K.~S. {Cheng}, R.~E. {Taam}, and W.~{Wang}, \emph{\apj} \textbf{617}, 480--489
  (2004).

\bibitem{2007arXiv0707.4279L}
X.-H. {Li}, F.-J. {Lu}, and Z.~{Li}, \emph{arXiv:0707.4279}  (2007).

\bibitem{2007arXiv0711.4107L}
D.~A. {Leahy}, and W.~{Tian}, \emph{arXiv:0711.4107}  (2007).

\bibitem{2003ApJ...591..361G}
E.~V. {Gotthelf}, \emph{\apj} \textbf{591}, 361--365 (2003).

\bibitem{2006MNRAS.368..497D}
S.~{Diehl}, and T.~S. {Statler}, \emph{\mnras} \textbf{368}, 497--510 (2006).

\bibitem{2000AAS...196.3704P}
G.~G. {Pavlov}, D.~{Sanwal}, G.~P. {Garmire}, et~al., \emph{BAAS} \textbf{32},
  733 (2000).

\bibitem{2007ApJ...656.1038D}
A.~A. {Deshpande}, and V.~{Radhakrishnan}, \emph{\apj} \textbf{656}, 1038--1043
  (2007).

\bibitem{2001MNRAS.328..393A}
A.~{Achterberg}, Y.~A. {Gallant}, J.~G. {Kirk}, et~al., \emph{\mnras}
  \textbf{328}, 393--408 (2001).

\bibitem{2006ApJ...653..325A}
E.~{Amato}, and J.~{Arons}, \emph{\apj} \textbf{653}, 325--338 (2006).

\bibitem{2008...Mori}
K.~{Mori}, O.~{Kargaltsev}, G.~{Pavlov}, et~al., \emph{Progr.\ Theor.\ Phys.\
  Supp., {\rm in press}}  (2008).

\bibitem{1995Natur.375...40M}
C.~B. {Markwardt}, and H.~{\"{O}gelman}, \emph{\nat} \textbf{375}, 40--42
  (1995).

\bibitem{1997ApJ...480L..13M}
C.~B. {Markwardt}, and H.~B. {\"{O}gelman}, \emph{\apjl} \textbf{480}, L13--L17
  (1997).

\bibitem{2006A&A...448L..43A}
F.~{Aharonian}, et~al., \emph{\aap} \textbf{448}, L43--L47 (2006).

\bibitem{2005Sci...307.1938A}
F.~{Aharonian}, et~al., \emph{Science} \textbf{307}, 1938--1942 (2005).

\bibitem{2006IAUJD...2E..53D}
O.~C. {de Jager}, \emph{On the Present and Future of Pulsar Astronomy, 26th
  meeting of the IAU, Joint Discussion 2, JD02, \#53}  (2006).

\bibitem{2007ApJ...670..643K}
O.~{Kargaltsev}, G.~G. {Pavlov}, and G.~P. {Garmire}, \emph{\apj} \textbf{670},
  643--654 (2007).

\bibitem{2007Ap&SS.309..197G}
Y.~A. {Gallant}, \emph{\apss} \textbf{309}, 197--202 (2007).

\bibitem{2007arXiv0709.3614C}
C.~{Chang}, A.~{Konopelko}, and W.~{Cui}, \emph{arXiv:0709.3614}  (2007).

\bibitem{2001ApJ...563..806B}
J.~M. {Blondin}, R.~A. {Chevalier}, and D.~M. {Frierson}, \emph{\apj}
  \textbf{563}, 806--815 (2001).

\bibitem{2007Ap&SS.309..189H}
D.~{Horns}, F.~{Aharonian}, A.~I.~D. {Hoffmann}, et~al., \emph{\apss}
  \textbf{309}, 189--195 (2007).

\bibitem{2007arXiv0708.1050A}
J.~{Arons}, \emph{arXiv:0708.1050}  (2007).

\end{thebibliography}


\end{document}